\documentclass{elsart}
\usepackage{graphics}
\usepackage{epsfig}
\usepackage{subfigure}
\usepackage{amssymb}
\usepackage{mathrsfs}
\def\shiftleft#1{#1\llap{#1\hskip 0.04em}}
\def\shiftdown#1{#1\llap{\lower.04ex\hbox{#1}}}
\def\thick#1{\shiftdown{\shiftleft{#1}}}
\def\b#1{\thick{\hbox{$#1$}}}
\newcommand{\cal}{\mathscr}

\begin{document}

\begin{frontmatter}



\title{Quasi-elastic knockout of pions and kaons from nucleons by
high-energy electrons and quark microscopy of ``soft'' meson
degrees of freedom in the nucleon}


\author{V.G. Neudatchin, I.T. Obukhovsky, L.L. Sviridova, 
N.P. Yudin}
\ead{obukh@nucl-th.sinp.msu.ru}
\address{Institute of Nuclear Physics, Moscow
State University,\\
119899 Moscow, Russia}
\begin{abstract}
Electro-production of pions and 
kaons at the kinematics of quasi-elastic knockout (which is well known
in the physics of atomic nucleus and corresponds to the
$t$-pole diagram) is proposed for obtaining their momentum distribution (MD)
in various channels of virtual decay $N \to B+\pi$, $B=N$, $\Delta$, $N^*$, 
$N^{**}$, and $N \to Y+K$, $Y=\Lambda$, $\Sigma$. It is a powerful tool for
investigation of a quark microscopic picture of the meson 
cloud in the nucleon. A model of scalar $q \bar{q}$ ($^3P_0$) fluctuation in 
the non-trivial QCD vacuum is used to calculate pion and kaon momentum 
distributions (MD) in these channels.
\end{abstract}

\end{frontmatter}

%\date{}
%\maketitle

\section{Introduction}

Investigation of structure of a composite system by means of
quasi-elastic knockout of its constituents has been playing a very
important role in the micro-physics~\cite{b1,b2,b3,b4,b5,b6}. In a broad 
sense, the term
"quasi-elastic knockout" means that a high-energy projectile (electron,
proton, etc.) instantaneously knocks out a constituent --- an electron from
an atom, molecule or solid film~\cite{b1,b2}, a nucleon~\cite{b3} or cluster 
\cite{b4} from a 
nucleus, a meson from a nucleon~\cite{b5} or nucleus~\cite{b6} --- 
transferring a high momentum in an "almost free" binary collision to the 
knocked-out particle and 
leading to controllable changes in the internal state of the target. Such
process corresponds to the Feynman pole diagram Fig.~\ref{f1}.

Exclusive quasi-elastic knockout coincidence experiments resolve
individual states of the final system (different channels of the virtual
decay of the initial composite system into a constituent and a final
system-spectator in a given excited state). First of all, such experiments,
in accordance with the laws of binary collisions~\cite{b2}, give the missing
momentum and energy, i.e. the momentum of the constituent and its
binding energy in the channel being considered. Of course, these values
should be much smaller than momentum and energy of the final
knocked-out particle. By varying kinematics, one can directly
measure the momentum distribution (MD) of a constituent in different
channels. For example, in the case of nuclei, such experiments make it
possible to determine the momentum distributions of nucleons in
different nuclear shells. Second, in such experiments, it is possible to
measure spectroscopic factors of constituent separation in various
channels. For nuclei, the spectroscopic factors determine probabilities
of exciting different states of the residual nucleus-spectator $A-1$ after
the knockout of a nucleon from the initial nucleus $A$ (it corresponds to
the spectrum of relevant fractional-parentage coefficients~\cite{b7}, used 
in the theory of many-particle nuclear shell model).

In the present paper, we discuss a problem of how the above-mentioned
experience can be extended to the investigation of a quark microscopic
picture of meson cloud in the nucleon. We investigate here a pion
production on nucleons by electrons with energy of a few GeV within the
kinematics of quasi-elastic knockout. We rely on both the corresponding
international literature~\cite{b8,b9} and our previous results~\cite{b5,b10}.

Our relativistic analysis is done within the instantaneous form of
dynamics in the initial proton rest frame (laboratory frame) as far as 
the kinematics criterion of the quasi-elastic character of the process (a 
large momentum of the knocked-out particle and a small recoil momentum of 
the residual system-spectator) is especially distinct here. If this
reference frame is accepted, a significant contribution from the pole
$Z$-diagram (Fig.~\ref{f1}b) should be taken into account, too. The 
corresponding
amplitude may easily be obtained from the amplitude of the diagram
Fig.\ref{f1}a using the crossing symmetry relations~\cite{b11}.

Within the quasi-elastic knockout kinematics (which includes the condition
$Q^2 \sim$ 1-3 (GeV/$c$)$^2$, $q^\mu q_\mu=-Q^2$ being the 4-momentum squared
of the virtual photon) the pole mechanisms of Fig.~\ref{f1}a and 
Fig.~\ref{f1}b dominate~\cite{b10}. The physical attractiveness of such 
kinematics in the 
quasifree collision can be illustrated, in particular, 
by the fact, that the problem of the gauge invariance of the knockout process 
is reduced here to that for the two-body electron-meson free collision, i.e. 
to the solved problem (see formal comments below and in 
Refs.~\cite{b5,b10}). 
The authors of a preceding pioneering papers~\cite{b8,b9} have analyzed 
an exclusive pion electro-production experiment $p(e,e' \pi^+)n$ 
\cite{b12,b12a,b12b}
with $Q^2$ values ranging from 0.7 to 3.3 (GeV/$c$)$^2$ within the
pole approximation and have reconstructed the MD of pions in the
channel $p \to n+ \pi^+$. But the problem of how much non-pole
diagrams are suppressed here and how to make this suppression maximal
by varying kinematics was not discussed.

In Refs.~\cite{b8,b9}, the light-front dynamics was used and the MDs were
expressed in terms of ($x$, $k_\perp$) variables. This approach may be
very helpful in more complicated cases (the interference of many
different diagrams) as far as the contributions of all $z$-diagrams
disappear here, but for our simplest situation of two pole amplitudes with
the common vertex function such approach is not indispensable.

Further, papers~\cite{b9,b10} contain an important note that while for pion 
photo-production reactions ($Q^2=0$) contributions of the $t$-pole and 
$s$-pole diagrams are compatible, for the pion electroproduction the 
relative contribution of non-$t$-pole diagrams decreases with increase of 
the $Q^2$ value. But as far as the sum of the $t$-pole and $s$-pole diagrams 
is maintained gauge invariant~\cite{b10}, the $t$-amplitude itself, being 
very predominant at $Q^2 \sim$ 1-3 (GeV/$c$)$^2$, becomes gauge invariant 
with a good accuracy. 

In the papers~\cite{b8,b12}, it was also pointed out that the 
Rosenbluth separation~\cite{b12} permits us to extract from the 
$p(e,e' \pi )n$ 
experiment at large enough $Q^2$ values not only the MD of pions in 
nucleon (analyzing the longitudinal cross-sections $d \sigma_L/dt$) but 
also the MD of rho-mesons in nucleon (analyzing the transverse cross-sections 
$d \sigma_T/dt$ and having in mind the non-diagonal subprocess 
$\rho^++ \gamma_T \to \pi^+$).

In the present paper, we follow this way, but, as a first step, consider 
only pionic and kaonic clouds in the nucleon. Namely, we extract from the 
$p(e,e' \pi )n$
experimental data on $d \sigma_L/dt$~\cite{b12,b12a,b12b} the pion MD 
$|\Psi_p^{n\pi}({\bf k})|^2$ and show, as a confirmation of our approach, that it is
very close to MD determined independently from a $\pi N$-potential 
\cite{b13,b14}, which had been reconstructed from $\pi N$ scattering data. 
But we do not restrict ourselves by this phenomenology
and also consider here, as a central point, a quark microscopic picture of
the pionic and kaonic clouds.

As it was mentioned above, the quasi-elastic knockout method is very
suitable for the investigation of microscopic properties of many-particle
systems. In the present paper, such system is exemplified by a nucleon
considered within the quark model of $^3P_0$ scalar fluctuation 
\cite{isg,st,ya,b15,sw}. 
In this model the nucleon is characterized by a $q^4 \bar{q}$ configuration 
with a meson as a virtual composite ($q \bar{q}$) particle. Conceptually, the situation 
is similar to that for virtual clusters in the atomic nuclei, where the
valuable opportunities of quasi-elastic knockout were discussed in details
\cite{b4} (e.g., the deexcitation of a virtual excited cluster in a nucleus 
by a proton blow in $(p,p \alpha)$ reaction).

The channels of the virtual decay $p \to B+ \pi$ ($B=N$, $\Delta$,
$N^*(1/2^-,3/2^-)$, $N^{**}(1/2^+,$ Roper)) and $p \to Y+K$ ($Y= \Lambda$,
$\Sigma$) are considered in our microscopic approach, the corresponding
MDs $|\Psi_p^{B \pi}({\bf k})|^2$, $|\Psi_p^{Y K}({\bf k})|^2$ and
spectroscopic factors $S_p^{B \pi}$, $S_p^{Y K}$ are calculated,
having in mind possible experiments in JLab. From the formal point of
view, the main problem here is the re-coupling of quark coordinates when
forming the virtual meson. In the shell-model theory of nucleon
clustering in light nuclei, this problem had been solved long ago by
joining the many-particle fractional-parentage technique with the
Talmi-Moshinsky-Smirnov (TMS) transformation of the oscillator wave
functions from a single nucleon coordinates to the cluster Jacobi 
coordinates~\cite{b16}.

It should be stressed, that all this physics of "soft" hadron degrees of
freedom in the nucleons and nuclei discussed in the present paper and
connected with the moderate $Q^2$ values 2-4 (GeV/$c$)$^2$ remains
beyond the interests of the scientific community. Its attention is 
concentrated on the frontier problem of quark asymptotic degrees 
of freedom~\cite{b17}, which corresponds to $Q^2=10-20$ (GeV/$c$)$^2$ and 
very small cross-sections.
The general aim of the present paper is to renew the interest in the
investigation of the hadron virtual components in nucleons and nuclei.

Our paper is organized as follows. In the second section, the relativistic
theory of quasi-elastic knockout reactions $p(e,e' \pi)n$ is briefly
presented, following Refs.~\cite{b5,b10}. The third section contains 
analysis of the basic formal problem of projecting the nucleon $q^4 \bar{q}$ 
wave function into the different $B+\pi$ and $Y+K$ channels within the 
microscopic $^3P_0$ model. The resulting momentum distributions 
and spectroscopic factors are discussed in the final fourth section.

%%%%%%%%%%%%%%%%%%%%%%%%%%%%%%%%%%%%%%%%%%%%%%%%%%%%%%%%%%%%%%%%%%%%%%%
\section{Elements of quasi-elastic knockout $p(e,e' M)B$ theory}

The general expression for the pion electro-production cross section is
well known~\cite{b12,b18}:

\begin{equation}
\frac{d^4 \sigma}{dW^2 dQ^2 dt d\phi_x}=
\Gamma
\left\{ \varepsilon \frac{d \sigma_L}{dt} + \frac{d \sigma_T}{dt}
+ \sqrt{2 \varepsilon  (1+\varepsilon )} \frac{d \sigma_{LT}}{dt}
cos \phi_x
+\varepsilon \frac{d \sigma_{TT}}{dt}cos 2 \phi_x
\right\},
\label{e1}
\end{equation}
where
$t = (p_R-p_T)^2=k^2$, $p_T=( E_T ,{\bf p}_T)$, $p_R=( E_R ,{\bf p}_R)$
are, respectively, 4-momenta of a target particle and of a recoil particle 
(baryon); $W^2=(p_x+p_R)^2$ is the invariant mass of final hadrons, 
$p_x=( E_x ,{\bf p}_x)$
being 4-momentum of a product particle (meson);
$Q^2=-q^2$, $q$ being the virtual-photon 4-momentum;
$\phi_x$ is the angle between the plane spanned
by the initial and final electron momenta $({\bf p}_e,{\bf p}^{\prime}_e)$ 
and the plane spanned by the $({\bf p}_x,{\bf p}_R)$ momenta; 
the quantity
\begin{equation}
\varepsilon = \left[ 1+ \frac{2 {\bf q}^2}{Q^2} tan^2 \frac{\theta_e}{2}
\right] ^{-1}
\label{e2}
\end{equation}
characterizes the degree of longitudinal polarization of the virtual photon, 
$\theta_e$ is electron's scattering angle, and
\begin{equation}
\Gamma=
\frac{\alpha}{(4 \pi )^2} \frac{W^2-M_T^2}{Q^2E_e^2M_T^2}
\frac{1}{1-\varepsilon},
\label{e4}
\end{equation}
$E_e$ is the initial-lepton (electron) energy, $M_T$ is the target mass, 
and $\alpha$ is the fine structure constant.
Finally, $d \sigma_{\scriptscriptstyle L}/dt$ is the longitudinal cross section,
$d \sigma_{\scriptscriptstyle T}/dt$ is the transverse one, and
$d \sigma_{\scriptscriptstyle LT}/dt$, $d \sigma_{\scriptscriptstyle TT}/dt$
represent the interference terms~\cite{b12}.

Experimental results are presented in terms of $d \sigma_i /dt$.
For the longitudinal virtual photons the electromagnetic vertex of the diagram Fig.~\ref{f1} is
characterized by a subprocess $\pi^++\gamma_L^* \to \pi^+$~\cite{b5,b9}.
All other subprocesses are suppressed at $Q^2 \ge 1$ (GeV/$c$)$^2$.
For the transverse photons  a non-diagonal 
subprocess $\rho^++\gamma_T^* \to \pi^+$ \cite{b5,b9} begins to dominate 
over $\pi^++\gamma_L^* \to \pi^+$ at large enough $Q^2 $
($Q^2 \ge 1$ (GeV/$c$)$^2$). So, the Rosenbluth separation~\cite{b12} of the 
cross section into longitudinal $d \sigma_{\scriptscriptstyle L}/dt$ and 
transverse 
$d \sigma_{\scriptscriptstyle T}/dt$ parts
permits us to investigate both the pionic structure of the nucleon and 
its $\rho$-meson structure by means of the same process of exclusive 
pion electroproduction. In the present paper, we consider, as a first step,
only the pionic and kaonic clouds (only the longitudinal cross sections).

According to the general rules of the field theory ~\cite{b19}, the wave
function of a pseudo-scalar constituent $x$ in the target $T$,
which corresponds to the pole diagram Fig.~\ref{f1}a, is defined as

\begin{equation}
\Psi_T^{Rx}({\bf k})=\frac{{\cal M}({\scriptstyle T \to Rx})}{E_T-E_R-E_x},
\label{e5}
\end{equation}
where ${\cal M}({\scriptstyle T \to Rx})$ is an amplitude of the process of 
virtual decay $T \to Rx$. 

The probability to find particle $x$ in the channel of virtual decay
$T \to Rx$ is characterized by the spectroscopic factor 
\begin{equation}
\int \overline{|\Psi_T^{Rx}({\bf k})|^2} d \tau = S_T^{Rx}
\label{e6}
\end{equation}
with the integration measure given by $d \tau =d^3{k} /[(4 \pi)^3E_TE_RE_x]$.
It is convenient to define the ``radial'' part $R_T^{Rx}$ of the wave function 
(\ref{e5}) as
\begin{equation}
\frac{\overline{|\Psi_T^{Rx}({\bf k})|^2}}{(2 \pi)^32E_T2E_R2E_x } =
\frac{1}{4 \pi}|R_T^{Rx}(k^2)|^2,
\label{e7}
\end{equation}
that is normalized to the spectroscopic factor~(\ref{e6}) by the equation
$\int\limits_0^\infty |R_T^{Rx}(k^2)|^2 k^2 dk = S_T^{Rx}$.

The formulas presented above are quite general: they are valid for atoms,
nuclei, and hadrons. We will now specify these formulas for the case of the
quasi-elastic knockout of pions from nucleons. So, $p_T=p$ is an initial
nucleon momentum; $p_R=p'$ is a final nucleon momentum; and
$p_x=k'$ is a final pion momentum.

In the reactions with longitudinal photons, a process with the virtual
pions $p \to n+\pi^+$ dominate over processes with other kinds
of virtual mesons. So, to obtain the pion wave function, we need
experimental data on the longitudinal cross section 
$d \sigma_{\scriptscriptstyle L}/dt$.

The longitudinal cross section may be expressed in terms of the wave
function of pion in nucleon
$\Psi_p^{n \pi}({\bf k})={\cal M}(p \to n+\pi^+)/(k_0-\omega_\pi 
({\bf k}))$ as~\cite{b5,b10}
\begin{equation}
\frac{d \sigma_{\scriptscriptstyle L}}{dt}=\frac{\alpha}{8W|{\bf q}^{cm}| (W^2-M_N^2)^2}
\overline{|\Psi_p^{n \pi}({\bf k})|^2}
\frac{(k_0-\omega_\pi ({\bf k}))^2}{(k^2-m_\pi^2)^2}
F^2_{\pi \pi \gamma}(Q^2)
\{ (k+k') \cdot e_{\lambda=0} \}^2,
\label{e10}
\end{equation}
where $\omega_\pi ({\bf k})=\sqrt{{\bf k}^2+m_\pi^2}$,
${\bf q}^{cm}$ is the virtual photon momentum in the c.m. frame of final 
particles, $e_{\lambda=0}$ is the photon polarization unit 4-vector for 
longitudinal photons,  and $\overline{|\Psi_p^{n \pi}({\bf k})|^2}$ is 
squared and averaged over spins wave function Eq.~(\ref{e5}) for pions.
$F_\pi(Q^2)$ is the electromagnetic form factor for the $\pi \pi \gamma$ 
vertex; it is accepted to be equal to the free pion form factor
\begin{equation}
F_{\pi \pi \gamma}(Q^2)=[1+(Q^2/0.54 (\mbox{GeV}/c)^2)]^{-1},
\label{e9}
\end{equation}
which corresponds to the value of charge pion radius 
$[<r_{\pi}^2>_{ch}]^{1/2}=$ 0.656 fm~\cite{gr} or to the value of the pion 
quark radius 
$b_\pi\simeq 0.3$ fm considered here following Ref.~\cite{isg1}~\footnote{In 
the constituent quark model the pion charge radius includes contributions
both quarks and the $\rho$-meson pole (the vector meson dominance model): 
$[<r_{\pi}^2>_{ch}]^{1/2}=[3b_{\pi}^2/4+6/m_{\rho}^2]^{1/2}\simeq$ 0.68 fm.}.
However, at large $Q^2\gtrsim 0.7- 1.$ 
Gev$^2/c^2$ we use more exact data on the charge pion e.m. form factor 
recently extracted from the longitudinal $p(e,e^{\prime}\pi^+)n$ cross 
section data \cite{vol}. We shall keep in mind this consistent description 
when discussing vertex constants below.

We can also present the longitudinal part of the cross section (1) within 
the pole approximation in the form which is commonly accepted in the physics 
of atomic nucleus and explicitly corresponds to the kinematics of coincidence 
experiment~\cite{b5}:
\begin{equation}
\frac{d^5 \sigma_L}{dE_e'd\Omega_e'd\Omega_\pi}=
|{\bf k}'|\,\omega_\pi({\bf k}')\, \overline{|\Psi_N^{B \pi}({\bf k})|^2}
\left( \frac{\omega_\pi({\bf k})}{k_0+ \omega_\pi({\bf k})}\right)^2
\left( 1-\frac{{\bf k}'}{E_\pi({\bf k}')cos\theta_\pi}\right)
\frac{d\sigma_{el}^{free}}{d\Omega_\pi},
\label{neud}
\end{equation}
where $\theta_\pi$ is the angle between momenta $\bf p_e'$ and $\bf k'$
and $ d\sigma_{el}^{free}/d\Omega_\pi $ is the cross section of free $e\pi$
scattering. However, we will work here with Eq.~(\ref{e10}), which is more
conventional in the physics of mesons.

Using the following formula for the amplitude of the virtual decay
$ p \to n+\pi^+$ 
\begin{equation}
{\cal M}(p \to n+\pi^+)=\sqrt{2}g_{\pi {\scriptscriptstyle NN}}
F_{\pi {\scriptscriptstyle NN}}(k^2) 
\bar{u}_{{\scriptscriptstyle N}^{\prime}}(p^{\prime}) 
i\gamma^5u_{\scriptscriptstyle N}(p)
\label{e11}
\end{equation}
($u_{\scriptscriptstyle N}(p)$ being the Dirac spinor of a nucleon
normalized on the nucleon mass, 
$\bar{u}_{\scriptscriptstyle N}u_{\scriptscriptstyle N}=
2M_{\scriptscriptstyle N}$), 
we can express the wave function (\ref{e5}) through the vertex function
$g_{\pi {\scriptscriptstyle NN}}F_{\pi {\scriptscriptstyle NN}}(k^2)$ in 
the strong $\pi NN$ vertex
\begin{equation}
\overline{|\Psi_p^{n \pi}({\bf k})|^2}=
2 g^2_{\pi {\scriptscriptstyle NN}}F_{\pi {\scriptscriptstyle NN}}^2(k^2) 
\frac{|k^2|}{(k_0- \omega_\pi ({\bf k}))^2}.
\label{e12}
\end{equation}
Here $k^2 \simeq -{\bf k}^2$. Often the form factor is parametrized in
the monopole form:
\begin{equation}
F_{\pi {\scriptscriptstyle NN}}(k^2)= 
\frac{\Lambda_{\pi}^2-m_{\pi}^2}{\Lambda_\pi^2-k^2}.
\label{e13}
\end{equation}
In this case the quasi-elastic knockout of pions can, in principle,
allow us to get the cut-off constant $\Lambda_{\pi}$.

It is also possible to determine the wave function of pion in nucleon
using a $\pi N$-potential, that was obtained from the $\pi N$ scattering
data. We have used a potential by I.R. Afnan~\cite{b13}, which includes a
pole and a contact terms:
\begin{equation}
V(k,k',E)= \frac{f_0(k)f_0(k')}{E-M_0}-h_0(k)h_0(k'),
\label{e14}
\end{equation}
where $M_0$ is a bare-nucleon mass. The functions $f_0(k)$ and $h_0(k)$
are choose in such a way as to obtain a satisfactory description of the
phase shifts for $\pi N$ scattering.

The wave function~(\ref{e6}) is determined from the residue of
the exact $\pi N$ propagator in the pole
\begin{equation}
G(k,k',E)=\frac{f(k,E)
f(k',E)}{E-M_{\scriptscriptstyle N}},
\label{e15}
\end{equation}
where $M_N$ is the physical nucleon mass.

Thus, we have for the `''radial'' part of wave function
\begin{equation}
R_p^{n \pi}(k)=
\frac{\sqrt{2}f(k,E=M_{\scriptscriptstyle N})} 
{M_{\scriptscriptstyle N} - \omega_\pi({k}) - E_0({k})},
\label{e16}
\end{equation}
where $E_0({k})=\sqrt{{\bf k}^2+M_0^2}$ [we use in 
Eqs.~(\ref{e14})-(\ref{e19}) the notation $k=|{\bf k}|$].

The functions $f(k,E)$ and the mass of the bare nucleon $M_0$ can be
found from the equations presented in Ref.~\cite{b13}:
\begin{equation}
f(k,E)=f_0(k,E)+h_0(k) \tau_0(E)
\int h_0(k')f_0(k')D_{\pi {\scriptscriptstyle N}}(k',E) k'^2dk',
\label{e17}
\end{equation}

\begin{equation}
D_{\pi {\scriptscriptstyle N}}(k,E)=[E- \omega_\pi({k}) - 
E_0({k})+i \delta]^{-1},
\label{e18}
\end{equation}

\begin{equation}
\tau_0(E)=- \left[ 1+ \int |h_0(k)|^2 
D_{\pi {\scriptscriptstyle N}}(k,E) k^2dk
\right]^{-1}.
\label{e19}
\end{equation}

%%%%%%%%%%%%%%%%%%%%%%%%%%%%%%%%%%%%%%%%%%%%%%%%%%%%%%%%%%%%%%%%%%%%%%%%%%%%%

\section{Quark microscopic picture of the $N \to B+\pi$ and $N \to
Y+K$ virtual channels within the model of $^3P_0$ scalar fluctuation}

The formal description of the quasi-elastic knockout of composite
particles (clusters) from atomic nuclei is a well-developed procedure 
\cite{b16}. For example,
in a channel of virtual decay $A_i \to (A-4)_f+\alpha_n$, the wave
function of mutual motion $(A-4)_f-\alpha_n$ can be defined as
\begin{equation}
\Psi_i^{f \alpha_n}({\bf R})=c <(A-4)_f \alpha_n |A_i>,
\label{e20}
\end{equation}
where $c$ is a constant factor. Nucleon numbers in the virtually excited
$\alpha$-particle are fixed. The integration is carried out over the internal
variables of the subsystems $(A-4)_f$ and $\alpha_n$. The technique of
fractional parentage coefficients is used along with the
Talmi-Moshinsky-Smirnov transformation of the oscillator wave
functions from a single nucleon coordinates to the cluster Jacobi
coordinates~\cite{b16}. In the quasi-elastic knockout process like 
$A_i(p,p^{\prime}\alpha_0)(A-4)_f$ with protons of 500-1000 MeV energy the
non-diagonal amplitudes $p+\alpha_n \to p+\alpha_0$ should be taken
into account \cite{b4}. The observable MD of the virtual $\alpha$-particles 
in the mentioned channel is, in fact, a squared sum of a few different
comparable components $\Psi_i^{f \alpha_n}({\bf q})$ taken for each
$n$ with its own amplitudes of $\alpha_n \to \alpha_0$ deexcitation, which
are calculated within the Glauber-Sitenko multiple scattering theory 
\cite{b4,b16}.
The MDs for various final states $f$ may differ greatly from each other.

The physical content of the ''microscopic'' hadron theory
corresponds, in general, to this concept. It is true, at least, for
QCD motivated quark models taking into account the $q\bar q$ pair creation, 
the flux-tube breaking model~\cite{isg,st} or 
merely the ``naive'' $^3P_0$ model~\cite{ya,b15}.
Namely, the nucleon as a three-quark system with a $q\bar q$ fluctuation 
(the $1234\bar 4$ system in Fig.~\ref{f2}) virtually decays into subsystems
124 and $3\bar 4$ which can be formed in various states of internal 
excitation. Only after the redistribution of quarks between two clusters 
$(123)+(4\bar 4)\to  (124)+ (3 \bar{4})$ the scalar $q \bar{q}$ fluctuation 
(color or colorless)\footnote{In the case of color $q\bar q$ fluctuation the 
necessity of redistribution is quite evident, and this leads to additional 
constraints beyond the OZI rule~\cite{sw}. Here we do not consider such 
effects and omit the color part of wave function.} 
becomes compatible with forming of the spin-less negative-parity pion and a 
baryon $B$. The formal method here is different from the formal method in 
the physics of nuclear clusters, although there are some common points: 
shell-model structure of $3q$-wave functions, fractional parentage 
coefficients, transformations of Jacobi coordinates, etc.

Note that the relation of the phenomenological $^3P_0$ models 
\cite{isg,st,ya,b15,sw} to the first principles of QCD has not been clearly 
established because of the essentially non-perturbative mechanism of 
low-energy meson emission. However, the models~\cite{isg,st,ya,b15,sw} have 
their good points: they satisfy the OZI rule and they enable to give 
reasonable predictions for transition amplitudes. 

The predictions which can be 
compared with the experimental data are of our main interest here. 
The most general prediction of the $^3P_0$ model is that the meson 
momentum distribution in the cloud should replicate the quark momentum 
distribution in the nucleon. For such a prediction the details of 
different $^3P_0$ models are not important, and we start here from an 
universal formulation proposed in Ref.~\cite{sw}. 
The $^3P_0$ production Hamiltonian is written in the covariant form as
a scalar source of $q\bar q$ pairs (the color part is omitted)
\begin{equation}
H_s=g_s\int d^3x\,\bar\psi_q(x)Z\psi_q(x)=g_s\int d^3x\,[\bar u(x)u(x)
+\bar d(x)d(x)+z\bar s(x)s(x)],
\label{hs}
\end{equation}
where $u(x)$, $d(x)$ and $s(x)$ are Dirac fields for the triplet of
constituent quarks and $Z$ is a diagonal 3$\times$3 matrix in the flavor
space 
\begin{equation}
\qquad\qquad\psi_q={\scriptscriptstyle\left(\!\!\!
\begin{tabular}{c}u\\[-8pt]d\\[-8pt]s\end{tabular}\!\!\!\right)},\qquad 
Z={\scriptscriptstyle\left(\!\!\!
\begin{tabular}{ccc}1&0&0\\[-8pt]0&1&0\\[-8pt]0&0&z\end{tabular}\!\!\!
\right).}
\label{z}
\end{equation}
The phenomenological parameter $z$ violating
the $SU(3)_{\scriptscriptstyle F}$ symmetry of the Hamiltonian (\ref{hs})
is required to reduce the intensity of creation of strange pairs in comparison 
with the non-strange pairs. Such reduction is necessary in any
variant of $^3P_0$ model because of a large difference between strange and 
non-strange quark masses that should violate the $SU(3)$ symmetry.

In terms of creation operators $b^\dagger_{{\bf p}\mu\alpha}$ and 
$d^\dagger_{{\bf p}\bar\mu\bar\alpha}$ defined in the Fock space
\begin{eqnarray}
\{b^\dagger_{{\bf p}^{\prime}\mu^{\prime}\alpha^{\prime}},
b_{{\bf p}\mu\alpha}\}&=&\delta_{\mu\mu^{\prime}}\,
\delta_{\alpha\alpha^{\prime}}\frac{E_p}{m_q}(2\pi)^3
\delta({\bf p}-{\bf p}^{\prime}),\qquad
b_{{\bf p}\mu\alpha}|0>=0, \quad etc.
%\nonumber\\
%\{d^\dagger_{{\bf p}^{\prime}\bar s^{\prime}
%\bar\alpha^{\prime}},
%d_{{\bf p}\bar s\bar\alpha}\}&=&\delta_{\bar s\bar s^{\prime}}\,
%\delta_{\bar\alpha\bar\alpha^{\prime}}\frac{E_p}{m_q}(2\pi)^3
%\delta({\bf p}-{\bf p}^{\prime}),\qquad
%d_{{\bf p}\bar s\bar\alpha}|0>
\label{bd}
\end{eqnarray}
the pair creation component of $H_s$ reads
\begin{equation}
H_{pair}=g_s\sum_{\alpha\mu}Z_{\alpha\alpha}\int\!
\frac{d^3p}{(2\pi)^3}\,\frac{m_q}{E_p}\int\!
\frac{d^3p^{\prime}}{(2\pi)^3}\,\frac{m_q}{E_{p^{\prime}}}
(2\pi)^3\delta({\bf p}\!+\!{\bf p}^{\prime})
\bar u({\bf p}\mu)v({\bf p}^{\prime}\bar\mu)
b^\dagger_{{\bf p}\mu\alpha}d^\dagger_{{\bf p}^{\prime}\bar\mu
\bar\alpha}\,,
\label{h}
\end{equation}
where $\alpha=u,d,s$, $E_p=\sqrt{m_q^2+{\bf p}^2}$, $m_q$ is the mass
of constituent quark $m_q\approx\frac{1}{3}M_N\approx\frac{1}{2}m_{\rho}$
 (or $m_q=m_s\approx\frac{1}{2}m_{\phi}$ in the strange sector), and the
standard normalization condition for quark bispinors is used,
$\bar{u}({\bf p}\mu)u({\bf p}\mu^{\prime})=\delta_{\mu\mu^{\prime}}$.
The intensity of pair production is determined by the value of 
phenomenological constant $g_s$ which is usually normalized on the 
amplitude of $N\to\pi+N$ transition, e.g. on the pseudo-vector coupling 
constant $f_{\pi NN}\backsimeq$ 1.0 (see below). The full Hamiltonian $H_s$
is considered here as an effective operator for description of the 
non-perturbative dynamics in terms of the production-absorption of $q\bar q$ 
pairs. 
In this formulation a particular mechanism of pair production is of a little 
importance. It could be the mechanism of flux-tube breaking~\cite{isg} or 
the Schwinger mechanism of pair production in a strong external field 
(see, e.g.~\cite{iz}) used in the ``naive '' $^3P_0$ model~\cite{ya}, etc.

Amplitudes of meson emission $N \to M+B$ and $M \to M_1+M_2$ are defined
as matrix elements of the Hamiltonian~(\ref{hs})
\begin{equation}
{\cal M}({\scriptstyle N \to M+B})=<\!M|\!<\!B| H_s|N\!>, 
\quad {\cal M}({\scriptstyle M \to M_1+M_2})=
<\!M_1|\!<\!M_2| H_s|M\!>,
\label{mnb}
\end{equation}
where the initial and final states are basis vectors of constituent
quark model (CQM). The non-relativistic shell-model states are commonly
used in calculations, but on the basis of covariant expression~(\ref{hs}) 
the relativistic Bethe-Salpeter amplitudes could be also defined.
In the non-relativistic approximation $\frac{E_p}{m_q}\approx$ 1
the wave function of $q\bar q$ fluctuation can be defined as
\begin{equation}
\gamma\psi_{q\bar q}({\bf r}_4,{\bf r}_{\bar 4})=
<q({\bf r}_4)|<\bar q({\bf r}_{\bar 4})| H_s|0>
\vert_{\frac{E_p}{m_q}\approx 1},
\label{qq}
\end{equation}
where the standard definitions of quark (anti-quark) Fock states are used
\begin{equation}
|q({\bf r})\!>=|q,\,\mu\,\alpha\,{\bf r}\!>=\int\frac{d^3p}{(2\pi)^3}
\frac{m_q}{E_p}\,e^{i{\bf p}\cdot{\bf r}}b^\dagger_{{\bf p}\mu\alpha}|0\!>,
\quad |q({\bf p})\!>=|q,\,\mu\,\alpha\,{\bf p}\!>=
b^\dagger_{{\bf p}\mu\alpha}|0\!>
\label{q}
\end{equation}
The factor $\gamma$ in the left-hand side of Eq.~(\ref{qq}) is a 
``coupling constant'' of the $^3P_0$ model for the non-strange $q\bar q$ 
pairs [$\alpha=u,d$ in Eq.~(\ref{h})]. 
A simple calculation leads to the explicit expression for both the wave 
function of $q\bar q$ fluctuation and the constant $\gamma$
\begin{eqnarray}
\psi_{q\bar q}({\bf r}_4,{\bf r}_{\bar 4})&=&
\int\frac{d^3p_4}{(2\pi)^3}\int\frac{d^3p_{\bar 4}}{(2\pi)^3}
(2\pi)^3\delta({\bf p}_4\!+\!{\bf p}_{\bar 4})\,e^{i({\bf p_4}\!-\!{\bf p}_{\bar 4})
\cdot{\b\rho}/2}\nonumber\\
&\times&\delta_{\alpha_4\bar\alpha_{\bar 4}}
<{\scriptstyle\frac{1}{2}}\,\bar\mu_{\bar 4}|{\b\sigma}
\cdot({\bf p_4}\!-\!{\bf p}_{\bar 4})
|{\scriptstyle\frac{1}{2}}\,\mu_{4}>,\qquad 
\gamma=Z_{\alpha\alpha}\frac{g_s}{2m_q}.
\label{wf}
\end{eqnarray}
This expression is usually used in the ``naive'' $^3P_0$ model. In the 
right-hand side
of Eq.~(\ref{wf}) the relative coordinates ${\b\rho}={\bf r}_{4}-
{\bf r}_{\bar 4}$ and ${\bf R}=({\bf r}_{4}+{\bf r}_{\bar 4})/2$ are used 
and a trivial factor 
$e^{i{\bf P}\cdot{\bf R}}=1$ (${\bf P}={\bf p}_4+{\bf p}_{\bar 4}=0$) is
omitted.

Using the explicit expression~(\ref{wf}) for the $q\bar q$ wave function
one can easily calculate the matrix elements~(\ref{mnb}) in the coordinate 
space with the standard technique of projecting the wave function~(\ref{wf}) 
onto the final meson-baryon states. The calculations are usually performed 
with an effective quark-meson vertex (the diagram in Fig.~\ref{f3}) 
\begin{equation}
H^{(3)}_{{\scriptscriptstyle M}qq}(\alpha,{\bf k})=<\!M,\,\alpha{\bf k}|
<\!q,\,t_{4}\mu_4{\bf p}_4| H_s|q,\,t_{3}\mu_3{\bf p}_3\!>,
\label{hm}
\end{equation}
where the meson state $|M,\,\alpha{\bf k}\!>$ is 
described with a simple (e.g. Gaussian) wave function
\begin{equation}
\Phi_{\scriptscriptstyle M}({\b\varkappa},{\bf k})=
\sqrt{2\omega_{\scriptscriptstyle M}({\bf k})}
(8\pi b_{\scriptscriptstyle M}^2)^{3/4}
e^{-{\b\varkappa}^2b_{\scriptscriptstyle M}^2},\quad {\b\varkappa}=
\frac{{\bf p}_3\!-\!{\bf p}_{\bar 4}}{2},\quad 
\omega_{\scriptscriptstyle M}({\bf k})=
\sqrt{m_{\scriptscriptstyle M}^2\!+\!{\bf k}^2}.
\label{mwf}
\end{equation}
For example, the pion $q\bar q$ state constructed on 
the basis of Gaussian wave function~(\ref{mwf}) has a form
\begin{eqnarray}
|\pi,\,\alpha{\bf k}\!>=i\int\frac{d^3p_3}{(2\pi)^3}\frac{m_q}{E_{p_3}}
\int\frac{d^3p_{\bar 4}}{(2\pi)^3}\frac{m_q}{E_{p_{\bar 4}}}
(2\pi)^3\delta({\bf k}\!-\!({\bf p}_3\!+\!{\bf p}_{\bar 4}))
\Phi_{\pi}(({\bf p}_3\!-\!{\bf p}_{\bar 4})/2,{\bf k})
\nonumber\\
\times\sum_{\mu_3\bar\mu_{\bar 4}}
(-1)^{1/2-\bar\mu_{\bar 4}}({\scriptstyle \frac{1}{2}}\mu_3
{\scriptstyle \frac{1}{2}}\!-\!\bar\mu_{\bar 4}|00)
\sum_{t_3t_{\bar 4}}(-1)^{1/2-\bar t_{\bar 4}}
({\scriptstyle \frac{1}{2}}t_3
{\scriptstyle \frac{1}{2}}\!-\!\bar t_{\bar 4}|1\alpha)
b^{\dagger}_{{\bf p}_3\mu_3t_3}
d^{\dagger}_{{\bf p}_{\bar 4}\bar\mu_{\bar 4}\bar t_{\bar 4}}|0\!>.
\label{pi}
\end{eqnarray}
The factor $\sqrt{2\omega_{\scriptscriptstyle M}}$ is introduced 
into the wave function~(\ref{mwf}) to ensure the normalization
\begin{equation}
<\!M,\,\alpha^{\prime}{\bf k}^{\prime}|M,\,\alpha{\bf k}\!>=
\delta_{\alpha\alpha^{\prime}}2\omega_{\scriptscriptstyle M}(\bf k)
(2\pi)^3\delta({\bf k}-{\bf k}^{\prime}),
\label{n}
\end{equation}
commonly used for bosons. In the first order of $\frac{v}{c}$ (i.e for 
small $k$,$p_i\lesssim m_q$) the $\pi qq$ vertex~(\ref{hm}) reads 
\begin{eqnarray}
H^{(3)}_{\pi qq(s)}(\alpha,{\bf k})&=&\frac{ig_s}{m_q}
(2\pi)^3\delta({\bf k}\!-\!({\bf p}_3\!-\!{\bf p}_4))\,
(2\pi b_{\pi}^2)^{3/4}\sqrt{\omega_{\pi}(\bf k)}\,\,
exp\left[-(({\bf p}_3\!+\!{\bf p}_4)/2)^2b_{\pi}^2\right]
\nonumber\\
&\times&<\!{\scriptstyle \frac{1}{2}}t_{4}|
{\tau_{\alpha}^{(3)}}^{\dagger}|{\scriptstyle \frac{1}{2}}t_{3}\!>
<\!{\scriptstyle \frac{1}{2}}\mu_{4}|
{\b\sigma}^{(3)}\cdot({\bf k}\!-\!({\bf p}_3\!+\!{\bf p}_{4}))|
{\scriptstyle \frac{1}{2}}\mu_{3}\!>+{\cal O}(\frac{v^2}{c^2}).
\label{npi}
\end{eqnarray}
It should be compared with the $\pi qq$ vertex for
the pseudo-vector (P.V.) coupling 
\begin{eqnarray}
H^{(3)}_{\pi qq({\scriptstyle P.V.})}(\alpha,{\bf k})
&=&i\frac{f_{\pi qq}}{m_{\pi}}
(2\pi)^3\delta({\bf k}\!-\!({\bf p}_3\!-\!{\bf p}_4))
<\!{\scriptstyle \frac{1}{2}}t_{4}|{\tau_{\alpha}^{(3)}}^{\dagger}
|{\scriptstyle \frac{1}{2}}t_{3}\!>\nonumber\\
&\times&<\!{\scriptstyle \frac{1}{2}}\mu_{4}|
{\b\sigma}^{(3)}\cdot({\bf k}-\frac{\omega_\pi(\bf k)}{2m_q}
({\bf p}_3\!+\!{\bf p}_{4}))|
{\scriptstyle \frac{1}{2}}\mu_{3}\!>+{\cal O}(\frac{v^2}{c^2})
\label{fpv}
\end{eqnarray}
usually used in ``chiral'' quark models (see, e.g. Ref.~\cite{br}). 
This vertex is defined as
\begin{eqnarray}
H^{(3)}_{\pi qq({\scriptstyle P.V.})}(\alpha,{\bf k})&=&
<\!\pi^\alpha({\bf k})|
<\!q,\,t_{4}\mu_4{\bf p}_4|H_{\scriptstyle P.V.}|q,\,t_{3}\mu_3{\bf p}_3\!>,
\label{pv}
\end{eqnarray}
with the P.V. interaction Hamiltonian for quarks
\begin{eqnarray}
H_{\scriptstyle P.V.}=-\,\frac{f_{\pi qq}}{m_{\pi}}\int d^3x
\bar\psi_q(x)\gamma^\mu\gamma^5\vec\tau\psi_q(x)\partial_\mu\vec\varphi_\pi(x),
\qquad |\pi^\alpha({\bf k})\!>=a^{\dagger}_{{\bf k}\alpha}|0\!>.
\label{hpv}
\end{eqnarray}
The spin-dependent  terms of Eqs.~(\ref{npi}) and (\ref{fpv}) are slightly 
different. This means that the recoil correction should be introduced in the 
$^3P_0$ vertex~(\ref{npi}) by the substitution\footnote{In the CQM 
the characteristic mass of a non-excited meson is about $2m_q$, 
but the mass of physical pion is too small $m_\pi\ll 2m_q$ (it is close to 
the zero mass of Goldstone boson), and thus a correction factor 
$\approx\frac{m_\pi}{2m_q}$ for the nucleon (baryon) recoil momentum would
be necessary to preserve the Galilean invariance of the CQM 
result (\ref{npi}) (see, e.g. Ref.~\cite{ob}, for details).}
\begin{equation}
{\b\sigma}^{(3)}\cdot({\bf k}\!-\!({\bf p}_3\!+\!{\bf p}_{4}))\quad\to\quad
{\b\sigma}^{(3)}\cdot({\bf k}-\frac{\omega_\pi(\bf k)}{2m_q}
({\bf p}_3\!+\!{\bf p}_{4})).
\label{cor}
\end{equation}

Another difference between Eqs.~(\ref{npi}) and (\ref{fpv}) steams from
the dependence of the $^3P_0$ vertex (\ref{npi}) on the pion wave function
(\ref{mwf}). As a result the $^3P_0$ vertex becomes a non-local operator 
which has a rather complicated form in the coordinate space (see the next 
subsection), 
but in the limit of the point-like pion $b_\pi\to0$ the standard (local) 
P.V. coupling comes from Eq.~(\ref{npi}). In this limit the non-local 
kernel in the coordinate space
\begin{equation}
\int\frac{d^3\varkappa}{(2\pi)^3}e^{i{\b\varkappa}\cdot({\bf r}_3-{\bf r}_4)}
e^{-{\b\varkappa}^2b_\pi^2}=(4\pi b_\pi^2)^{-3/2}
exp\left[-\frac{({\bf r}_3-{\bf r}_4)^2}{4b_\pi^2}\right]
\label{ker}
\end{equation}
approaches to the $\delta$-function $\delta({\bf r}_3-{\bf r}_4)$, and
the constant $g_s$ (and the constant $\gamma$ as well) becomes 
proportional to the P.V. $\pi qq$ coupling constant $f_{\pi qq}$:
\begin{equation}
\gamma=\frac{g_s}{2m_q}\to\frac{f_{\pi qq}}{2(2\pi b_\pi^2m_\pi^2)^{3/4}}
\label{nor}
\end{equation}
But such a relation of $g_s$ to the $\pi qq$ coupling constant $f_{\pi qq}$ 
is not convenient in use because of a singularity $\sim b_{\pi}^{-1}$ in the
limit $b_{\pi}\to 0$, and thus we must directly relate the $g_s$ to an 
observable (i.e. hadron) value, e.g. to the $\pi NN$ coupling constant 
$f_{\pi {\scriptscriptstyle NN}}$.
However, the relation between $f_{\pi NN}$ and $f_{\pi qq}$ depends on the 
CQM matrix element of $\pi qq$ vertex [(\ref{npi}) or (\ref{fpv})] (see the 
next subsection).  
%%%%%%%%%%%%%%%%%%%%%%%%%%%%%%%%%%%%%%%%%%%%%%%%%%%%%%%%%%%%%%%%%%%%%%%%
\subsection{$N \to B+ \pi$ channels}

In the model with the scalar Hamiltonian~(\ref{hs})
the amplitude of virtual decay $N \to B+ \pi^\alpha$, $\alpha =0, \pm 1$
is defined as
\begin{equation}
{\cal M}({\scriptstyle N \to B+ \pi^\alpha})=< B \pi^\alpha| H_s|N>
=3<B|H^{(3)}_{\pi qq(s)}(\alpha,{\bf k})|N>.
\label{e22}
\end{equation}
Remember that $ H_s|N>$ characterizes  the quark system $(1234 \bar{4})$, 
$B$ - the subsystem (124) and $\pi$ - the subsystem $(3\bar 4)$ 
(Fig.~\ref{f2}).
The factor 3 in the right-hand side reflects the identity of quarks.

The operator $H^{(3)}_{\pi qq(s)}(\alpha,{\bf k})$ 
has the following kernel in the coordinate space of $3q$-system\footnote{For 
the states with fixed initial and final nucleon (baryon) momenta 
${\bf P}={\bf p}_1\!+\!{\bf p}_2\!+\!{\bf p}_3$ and 
${\bf P}^{\prime}={\bf p}_1\!+\!{\bf p}_2\!+\!{\bf p}_4$ the last 
differential operator in Eq.(\ref{e24}) has the eigenvalue 
$\frac{2}{i}{\b\nabla}_{\scriptscriptstyle R}-{\bf k}=
{\bf P}\!+\!{\bf P}^{\prime}$.}:
\begin{eqnarray}
H^{(3)}_{\pi qq(s)}({\b\rho}^{\prime}_1{\b\rho}^{\prime}_2 {\bf R}^{\prime}, 
{\b\rho}_1{\b\rho}_2{\bf R};\,\alpha,{\bf k})=
\frac{ig_s}{m_{\pi}m_q}(2\pi b_{\pi}^2m_{\pi}^2)^{3/4}
\sqrt{\frac{\omega_{\pi}(\bf k)}{m_{\pi}}}{\tau^{(3)}_{\alpha}}^{\dagger}
\,\delta({\b\rho}^{\prime}_1\!-\!{\b\rho}_1)\,
e^{-i{\bf k}\cdot{\bf R}^{\prime}}
\nonumber\\
\times exp\left(i\frac{2}{3} {\bf k}\cdot{\b\rho}^{\prime}_2  \right)\, 
\hat{O}({\b\rho}^{\prime}_2,{\b\rho}_2;{\bf k})\,\,{\b\sigma}^{(3)}\!\cdot\!
\left[{\bf k}+\frac{\omega_\pi(\bf k)}{2m_q}
\left( \frac{2}{i} {\b\nabla}_{\rho_2}+\frac{2}{3}{\bf k}-
\frac{1}{3}(\frac{2}{i}{\b\nabla}_{\scriptscriptstyle R}-{\bf k})\right)
\right]
\label{e24}
\end{eqnarray}
Here
${\b\rho}_1={\bf r}_1-{\bf r}_2$,
${\b\rho}_2=({\bf r}_1+{\bf r}_2)/2-{\bf r}_3$,
${\b\rho}^{\prime}_2=({\bf r}_1+{\bf r}_2)/2-{\bf r}_4$, where
${\bf r}_i$ is the coordinate of $i$-th quark,
${\bf k}={\bf P}-{\bf P}^{\prime}={\bf p}_3-{\bf p}_4$ is a virtual 
pion momentum; 
$\sigma^{(3)}$ and $\tau^{(3)}$ are spin and isospin Pauli matrices 
for the third quark, $\tau^{(3)}_{\alpha =0, \pm 1}$ are spherical 
components of the vector $\vec\tau^{(3)}$ corresponding to the pion 
$\pi^\alpha$, $\vec\tau^{(3)}\cdot\vec\pi=\sum_{\alpha}
{\tau^{(3)}_{\alpha}}^{\dagger}\pi^{\alpha}$; $m_q=313$ MeV is the 
constituent quark mass. The pion energy on mass shell is
$\omega_\pi ({\bf k })=\sqrt{{\bf k}^2+m_\pi^2}$, but for virtual pions
the value $\omega_\pi ({\bf k })$ is defined by energy conservation in 
the vertex: $\omega_\pi ({\bf k })=M_{\scriptscriptstyle N}-\sqrt{{\bf k}^2+
M_{\scriptscriptstyle B}^2}$, where $M_{\scriptscriptstyle B}$ is the
mass of the baryon-spectator in the final state.
The nonlocal kernel 
$\hat{O}({\b\rho}^{\prime}_2, {\b\rho}_2)$ reads
\begin{equation}
\hat{O}({\b\rho}^{\prime}_2, {\b\rho}_2;{\bf k})=
exp \left( i \frac{1}{2} {\bf k}\cdot({\b\rho}_2\!-\!{\b\rho}^{\prime}_2)  
\right)(4 \pi b^2_\pi)^{\!-\!3/2}
exp \left(-\frac{({\b\rho}^{\prime}_2\!-\!{\b\rho}_2)^2}{4b_\pi^2}\right).
\label{e25}
\end{equation}
It includes as a factor the wave function of pion  
which is chosen in the Gaussian form~(\ref{mwf}) with 
$b_{\scriptscriptstyle M}$ being the pion radius $b_\pi$. The normalization
of the operator $\hat O$ is chosen so that in the limit
$b_{\pi}\to$ 0 the kernel~(\ref{e25}) approaches to the $\delta$-function
\begin{equation}
\lim_{b_{\pi}\to 0}\hat{O}({\b\rho}^{\prime}_2, {\b\rho}_2;{\bf k})=
\delta({\b\rho}^{\prime}_2-{\b\rho}_2).
\label{delt}
\end{equation}

In Eq.~(\ref{mnb}) $|N>$ and $|B>$ mean the internal wave functions of 
baryons~\cite{b20}
\begin{eqnarray}
|N(940)>&=&|s^3[3]_{\scriptscriptstyle X}L\!=\!0>_{\scriptscriptstyle TISM}
|[1^3]_{\scriptscriptstyle C}([21]_{\scriptscriptstyle S} 
\circ [21]_{\scriptscriptstyle T})
[3]_{ST}:[1^3]_{\scriptscriptstyle CST}>,\nonumber\\
|\Delta(1232)>&=&|s^3[3]_{\scriptscriptstyle X}
L\!=\!0>_{\scriptscriptstyle TISM}
|[1^3]_{\scriptscriptstyle C}([3]_{\scriptscriptstyle S} 
\circ [3]_{\scriptscriptstyle T})
[3]_{\scriptscriptstyle ST}:[1^3]_{\scriptscriptstyle CST}>,\nonumber\\
|N^*(1535)>&=&\{|s^2p[21]_{\scriptscriptstyle X}
L\!=\!1>_{\scriptscriptstyle TISM}
|[1^3]_{\scriptscriptstyle C}([21]_{S} \circ
[21]_{\scriptscriptstyle T})[21]_{\scriptscriptstyle ST}:
[21]_{\scriptscriptstyle CST}>\}^{J=1/2},
\nonumber\\
|N^{**}(1440)>&=&|sp^2[3]_{\scriptscriptstyle X}
L\!=\!0>_{\scriptscriptstyle TISM}
|[1^3]_{\scriptscriptstyle C}([21]_{\scriptscriptstyle S} 
\circ [21]_{\scriptscriptstyle T})[3]_{\scriptscriptstyle ST}:
[1^3]_{\scriptscriptstyle CST}>.
\label{e26}
\end{eqnarray}
The Young tableaux $[f]$ appear here in various subspaces 
(C, S, T, ST, CST); 
subscript TISM means ''the transitionally invariant shell model'', 
$S$ and $T$ values are defined unambiguously by 
$[f]_{\scriptscriptstyle S}$ and $[f]_{\scriptscriptstyle T}$ signatures, 
and ${\bf J}={\bf S}+{\bf L}$.
The radial parts of the baryon wave functions corresponding to a 
definite permutation symmetry $[f]_{\scriptscriptstyle X}$ 
are chosen as the harmonic oscillator (h.o.) wave 
functions (it makes easier the rearrangement of the quark coordinates): 

\begin{eqnarray}
|s^3[3]_{\scriptscriptstyle X}L\!=\!0\!>_{\scriptscriptstyle TISM}&=&
|0s({\b\rho}_1/\beta_1)> |0s({\b\rho}_2/\beta_2)\!>,
\nonumber\\
|s^2p[21]_{\scriptscriptstyle X}([2]\!\times\![1])
L\!=\!1\!>_{\scriptscriptstyle TISM}&=&
|0s({\b\rho}_1/\beta_1)\!> |1p({\b\rho}_2/\beta_2,m)\!>,\nonumber\\
|s^2p[21]_{\scriptscriptstyle X}([1^2]\!\times\![1])
L\!=\!1\!>_{\scriptscriptstyle TISM}&=&
|1p({\b\rho}_1/\beta_1,m)> |0s({\b\rho}_2/\beta_2)\!>,
\nonumber\\
|sp^2[3]_{\scriptscriptstyle X}L\!=\!0\!>_{\scriptscriptstyle TISM}
&=&\left(|2s({\b\rho}_1/\beta_1)\!>|0s({\b\rho}_2/\beta_2)\!>\right.
\nonumber\\
&+&\left.|0s({\b\rho}_1/\beta_1)\!> |2s({\b\rho}_2/\beta_2)\!>\right)
/\sqrt{2},
\label{e27}
\end{eqnarray}
where $|0s({\b\rho}/\beta)\!>=(\pi\beta^2)^{-3/4}
e^{-\rho^2/(2\beta^2)}\dots$, etc. are the h.o. basis states with 
$\beta_1=\sqrt{2}\,b$, $\beta_2=\sqrt{3/2}\, b$ 
($b \simeq 0.5-0.6$ fm is a nucleon radius in the CQM).
The $CST$ parts of the wave functions are written in the form 
of fractional parentage expansions, corresponding to the separation of 
the third quark.

For calculation of the $N\to\pi+N$ transition amplitude (the $\pi NN$ vertex)
we use the following explicit expression for the nucleon state
\begin{eqnarray}
|N,\mu t {\bf P}\!>&=&\int\frac{d^3p_1}{(2\pi)^3}\frac{m_q}{E_{p_1}}
\int\frac{d^3p_2}{(2\pi)^3}\frac{m_q}{E_{p_2}}
\int\frac{d^3p_3}{(2\pi)^3}\frac{m_q}{E_{p_3}}
(2\pi)^3\delta({\bf P}\!-\!\sum_{i}{\bf p_i})
\Phi_N({\b\varkappa}_1,{\b\varkappa}_2;{\bf P})
\nonumber\\
&&\times\sum_{\mu_i}\sum_{t_i}\sqrt{\frac{1}{2}}
\left[({\scriptstyle\frac{1}{2}}\mu_1
{\scriptstyle\frac{1}{2}}\mu_2|1\mu_{12})
(1\mu_{12}{\scriptstyle\frac{1}{2}}\mu_3|{\scriptstyle\frac{1}{2}}\mu)
({\scriptstyle\frac{1}{2}}t_1{\scriptstyle\frac{1}{2}}t_2|1t_{12})
(1t_{12}{\scriptstyle\frac{1}{2}}t_3|{\scriptstyle\frac{1}{2}}t)
\right.
\nonumber\\
&&+\left.\delta_{\mu_3\mu}\delta_{t_3t}
({\scriptstyle\frac{1}{2}}\mu_1{\scriptstyle\frac{1}{2}}\mu_2|00)
({\scriptstyle\frac{1}{2}}t_1{\scriptstyle\frac{1}{2}}t_2|00)
\right]
b^{\dagger}_{p_1\mu_1t_1}b^{\dagger}_{p_2\mu_2t_2}
b^{\dagger}_{p_3\mu_3t_3}|0\!>,
\label{np}
\end{eqnarray}
(a trivial color part is omitted). The nucleon wave function $\Phi_N$ 
is a product of Fourier transformations of TISM states~(\ref{e27}) 
\begin{eqnarray}
\Phi_N({\b\varkappa}_1,{\b\varkappa}_2;{\bf P})&=&
\sqrt{2E_N({\bf P})}\Phi_{h.o.}^{(12)}({\b\varkappa}_1)
\Phi_{h.o.}^{(3)}({\b\varkappa}_2),\quad E_N({\bf P})\!=\!\sqrt{M_N^2\!+\! 
{\bf P}^2},\nonumber\\
\Phi_{h.o.}^{(12)}({\b\varkappa}_1)&=&\int e^{-i{\b\varkappa}_1\cdot\rho_1}
|0s(\rho_1/\beta_1)\!>,\quad
\Phi_{h.o.}^{(3)}({\b\varkappa}_2)=\int e^{-i{\b\varkappa}_2\cdot\rho_2}
|0s(\rho_2/\beta_2)\!>,
\label{nho}
\end{eqnarray}
where the factor $\sqrt{2E_N({\bf P})}$ has been added to ensure the 
normalization
\begin{equation}
<\!N,\mu^{\prime}t^{\prime}{\bf P}^{\prime}|N,\mu t {\bf P}\!>=
\delta_{\mu^{\prime}\mu}\delta_{t^{\prime}t}2E_N({\bf P})
(2\pi)^3\delta({\bf P}\!-\!{\bf P}^{\prime})
\label{norm}
\end{equation}
commonly supposed for fermions.
The calculation of matrix elements 
\begin{eqnarray}
\cal M_s({\scriptstyle N}\!\to\!\pi\!+\!{\scriptstyle N})
&=&<\!\pi(q\bar q),\alpha {\bf k}|<\!N(3q)|
 H_s|N(3q)\!>=3<\!N|H^{(3)}_{\pi qq(s)}|N\!>,
\nonumber\\
\cal M_{\scriptscriptstyle P.V.} ({\scriptstyle N}\!\to\!\pi\!+\!{\scriptstyle N})
&=&<\!\pi^{\alpha}({\bf k})|<\!N(3q)|H_{\scriptscriptstyle P.V.}|
N(3q)\!>=3<\!N|H^{(3)}_{\pi qq({\scriptscriptstyle P.V.})}|N\!>
\label{mn1}
\end{eqnarray}
for both Hamiltonians~(\ref{hs}) and (\ref{fpv}) in the first order of 
$v/c$ [i.e. in the same approximation 
as in the case of Eqs.~(\ref{npi}) and (\ref{fpv})] leads to the expressions:
\begin{eqnarray}
\cal M_s({\scriptstyle N}\!\to\!\pi\!+\!{\scriptstyle N})&=&
\frac{5}{3}\frac{ig_s}{m_{\pi} m_q}(2\pi b^2_{\pi}m^2_{\pi})^{3/4}
\left[1-\frac{y_{\pi}}{3}\varphi_{\scriptscriptstyle N}(0)\right]
(1-y_{\pi})^{3/2}F^{(s)}_{\pi {\scriptscriptstyle NN}}({\bf k}^2)
\nonumber\\
&\times&{\tau^{(\scriptscriptstyle N)}_{\alpha}}^{\dagger}
{\b\sigma}^{(\scriptscriptstyle N)}\cdot\left[{\bf k}
-\frac{\omega_{\pi}(\bf k)}{2M_{\scriptscriptstyle N}}
({\bf P}\!+\!{\bf P}^{\prime})\right],
\label{mns}
\end{eqnarray}
\begin{eqnarray}
\cal M_{\scriptscriptstyle P.V.}({\scriptstyle N}\!\to\!\pi\!+\!
{\scriptstyle N})=
\frac{5}{3}\frac{if_{\pi qq}}{m_{\pi}}
F^{(\scriptscriptstyle P.V.)}_{\pi {\scriptscriptstyle NN}}({\bf k}^2)
{\tau^{(\scriptscriptstyle N)}_{\alpha}}^{\dagger}
{\b\sigma}^{(\scriptscriptstyle N)}\cdot\left[{\bf k}
-\frac{\omega_{\pi}(\bf k)}{2M_{\scriptscriptstyle N}}
({\bf P}\!+\!{\bf P}^{\prime})
\right],
\label{mnp}
\end{eqnarray}
where we have used the equality $M_{\scriptscriptstyle N}=3m_{q}$.
Here and further we use the notations
\begin{eqnarray}
y_{\pi}=\frac{2}{3}x_{\pi}^2(1+2x_{\pi}^2/3)^{-1},\quad
x_{\pi}=b_{\pi}/b,\quad
\varphi_{\scriptscriptstyle B}({\bf k})=3\omega_{\pi}({\bf k})
\left[M_{\scriptscriptstyle N}+M_{\scriptscriptstyle B}+
\omega_{\pi}({\bf k})\right]^{-1}
\label{ypi}
\end{eqnarray}
to simplify expressions.
The strong $\pi NN$ form factor $F_{\pi {\scriptscriptstyle NN}}$ in 
both models has a Gaussian form
\begin{eqnarray}
F^{(s)}_{\pi {\scriptscriptstyle NN}}({\bf k}^2)&=&
\left[1-y_{\pi}\varphi_{\scriptscriptstyle N}(0)/3\right]^{-1}
\left[1-y_{\pi}\varphi_{\scriptscriptstyle N}({\bf k})/3\right]
exp\left[-{\bf k}^2b^2(1+y_{\pi}/4)/6\right],\nonumber\\
F^{(\scriptscriptstyle P.V.)}_{\pi {\scriptscriptstyle NN}}({\bf k}^2)&=&
exp\left(-{\bf k}^2b^2/6\right) ,
\label{ff}
\end{eqnarray}
which is characteristic of the h.o. wave functions. Eqs.~(\ref{mns}) 
and (\ref{mnp})
should be compared with the standard definition of the P.V. vertex
for point-like nucleons  to 
obtain the normalization condition for $g_s$ and $f_{\pi qq}$:
\begin{eqnarray}
f_{\pi NN}&=&\frac{5}{3}f_{\pi qq}=\frac{5}{3}\frac{g_s}{m_q}
(2\pi b^2_{\pi}m^2_{\pi})^{3/4}\left[1-y_{\pi}
\varphi_{\scriptscriptstyle N}(0)/3\right](1-y_{\pi})^{3/2},\nonumber\\
\quad g_{\pi{\scriptscriptstyle NN}}&=&
\frac{2M_{\scriptscriptstyle N}}{m_{\pi}}f_{\pi{\scriptscriptstyle NN}},
\label{fg}
\end{eqnarray}
which relates the phenomenological constant of $^3P_0$ model $g_s$ to the 
$\pi NN$ coupling constant $g_{\pi{\scriptscriptstyle NN}}$. This relation
between $g_s$ and $g_{\pi{\scriptscriptstyle NN}}$ is more convenient than
Eq.~(\ref{nor}) as it does not require to consider a singular limit 
$b_{\pi}\to 0$. 

Starting from the value of $g_s$ fixed by Eq.~(\ref{fg}) we have calculated 
amplitudes for all the transitions $N\to\pi^{\alpha}+B$ and  
$N\to K^{\alpha}+Y$ with the same technique. For example, for baryons with
the wave functions defined by Eqs.~(\ref{e26}) and (\ref{e27}) the transition
amplitudes $N\to\pi^{\alpha}+B$ have the forms:
\begin{eqnarray}
{\cal M}_s({\scriptstyle N}\!\to\!\pi^{\alpha}\!+
\!{\scriptstyle \Delta})&=&
i g_{\pi {\scriptstyle N\Delta}}
F_{\pi {\scriptscriptstyle N\Delta}}({\bf k}^2)
{T^{(\scriptscriptstyle N\Delta)}_{\alpha}}^{\dagger}
{\bf{\scriptstyle \sum}}^{(\scriptscriptstyle N\Delta)}\!\cdot\!\!
\left[{\bf k}-\frac{\omega_{\pi}(\bf k)}{M_{\scriptscriptstyle N}\!
+\!M_{\scriptscriptstyle\Delta}}
({\bf P}\!+\!{\bf P}^{\prime})
\right],
\nonumber\\
{\cal M}_s({\scriptstyle N}\!\to\!\pi^{\alpha}\!+
\!{\scriptstyle N}_{{1/2}^-})&=&
g_{\pi {\scriptstyle N{N^{*}}}}
F_{\pi {\scriptscriptstyle N{N^{*}}}}({\bf k}^2)
{\tau^{(\scriptscriptstyle N)}_{\alpha}}^{\dagger}
\delta_{\mu\mu^{\prime}}\omega_{\pi}(\bf k),
\nonumber\\
{\cal M}_s({\scriptstyle N}\!\to\!\pi^{\alpha}\!+
\!{\scriptstyle N}_{{1/2}^+})&=&
ig_{\pi {\scriptstyle N{N^{**}}}}
F_{\pi {\scriptscriptstyle N{N^{**}}}}({\bf k}^2)
{\tau^{(\scriptscriptstyle N)}_{\alpha}}^{\dagger}
{\b\sigma}^{(\scriptscriptstyle N)}\!\cdot\!\!\left[{\bf k}
-\frac{\omega_{\pi}(\bf k)}{M_{\scriptscriptstyle N}\!
+\!M_{\scriptscriptstyle {N^{**}}}}
({\bf P}\!+\!{\bf P}^{\prime})
\right],
\label{ampl}
\end{eqnarray}
where the coupling constants and form factors are defined by 
expressions (see Ref.~\cite{ob} for details):
\begin{eqnarray}
f_{\pi{\scriptscriptstyle N\Delta}}&=&2\sqrt{2}\,\frac{g_s}{m_q}
(2\pi b^2_{\pi}m^2_{\pi})^{3/4}
\left[1-\frac{y_{\pi}}{3}\varphi_{\scriptscriptstyle \Delta}(0)\right]
(1-y_{\pi})^{3/2},
\quad g_{\pi{\scriptscriptstyle N\Delta}}\!=\!
\frac{M_{\scriptscriptstyle N}\!+\!M_{\scriptscriptstyle\Delta}}{m_{\pi}}
f_{\pi{\scriptscriptstyle N\Delta}},
\nonumber\\
f_{\pi {\scriptstyle N{N^{*}}}}&=&\frac{4}{3}\,
\frac{1}{2m_q b}\,\frac{ig_s}{m_q}
(2\pi b^2_{\pi}m^2_{\pi})^{3/4}(1-y_{\pi})^{5/2},
\quad g_{\pi{\scriptscriptstyle NN^{*}}}\!=\!
\frac{M_{\scriptscriptstyle N}\!+\!M_{\scriptscriptstyle N^{*}}}{m_{\pi}}
f_{\pi{\scriptscriptstyle NN^{*}}}
\nonumber\\
f_{\pi{\scriptscriptstyle NN^{**}}}&=&\frac{10}{9\sqrt{3}}\,
\frac{g_s}{m_q}(2\pi b^2_{\pi}m^2_{\pi})^{3/4}P_{\scriptscriptstyle R}(0)
(1-y_{\pi})^{3/2},\quad
g_{\pi{\scriptscriptstyle NN^{**}}}=\frac{M_{\scriptscriptstyle N}\!+
\!M_{\scriptscriptstyle N^{**}}}{m_{\pi}}
f_{\pi{\scriptscriptstyle NN^{**}}}, 
\label{const}
\end{eqnarray}
and
\begin{eqnarray}
F_{\pi{\scriptscriptstyle N\Delta}}({\bf k}^2)&=&
\left[1-y_{\pi}\varphi_{\scriptscriptstyle \Delta}(0)/3\right]^{-1}
\left[1-y_{\pi}\varphi_{\scriptscriptstyle \Delta}({\bf k})/3\right]
exp\left[-{\bf k}^2b^2(1+y_{\pi}/4)/6\right],\nonumber\\
F_{\pi {\scriptscriptstyle NN}^*}({\bf k}^2)&=&
\left\{1+\frac{{\bf k}^2b^2}{6}\left(1+\frac{y_{\pi}}{2(1-y_{\pi})}\right)
\left[\varphi^{-1}_{\scriptscriptstyle N^{*}}({\bf k})+\frac{y_{\pi}}{3}
\right]\right\}exp\left[-{\bf k}^2b^2(1+y_{\pi}/4)/6\right],\nonumber\\
F_{\pi{\scriptscriptstyle NN}^{**}}({\bf k}^2)&=&
P_{\scriptscriptstyle R}(0)^{-1}P_{\scriptscriptstyle R}({\bf k})
exp\left[-{\bf k}^2b^2(1+y_{\pi}/4)/6\right].
\label{nff}
\end{eqnarray}
with the following polynomial factor
\begin{eqnarray}
P_{\scriptscriptstyle R}({\bf k})=
\left\{\varphi_{\scriptscriptstyle N^{**}}({\bf k})-
\frac{9}{4}y_{\pi}\left[1+\frac{2}{3}(1-\frac{5}{6}y_{\pi})
\varphi_{\scriptscriptstyle N^{**}}({\bf k})\right]\right.\nonumber\\
+\left.\frac{{\bf k}^2b^2}{4}\left[1-y_{\pi}\left(1-\frac{y_{\pi}}{4}-
\frac{\varphi_{\scriptscriptstyle N^{**}}({\bf k})}{9}
(1-y_{\pi}+\frac{3}{4}y^2_{\pi})\right)\right]\right\},
\label{nffr}
\end{eqnarray}
for the Roper resonance.

Eqs.~(\ref{ampl})-(\ref{nffr}) implies that in the recoil term in 
Eq.~(\ref{cor}) we substitute an average value for the constituent quark 
mass $m_q=\frac{1}{6}(M_{\scriptscriptstyle B}+M_{\scriptscriptstyle N})$,
where $M_{\scriptscriptstyle B}$ is the mass of the final baryon $B=N,\,
\Delta,\,N^{*}=N_{1/2^-}(1535)$, $N^{**}=N_{1/2^+}(1440)$ in the vertex 
$N\to\pi+B$, and for $\omega_{\pi}({\bf k})$ is used the value
$M_{\scriptscriptstyle N}-\sqrt{M_{\scriptscriptstyle B}^2+{\bf k}^2}$,
which follows from the energy conservation in the vertex. 

The above formalism concerns the microscopic picture of Fig.~\ref{f2}, which 
corresponds to the diagram Fig.~\ref{f1}a. The diagram Fig.~\ref{f1}b 
represents creation of a virtual 
pair $\pi^+\pi^-$ and a virtual capture $N+\pi \to B$ instead of the virtual 
decay $N \to B+\pi$. The corresponding matrix element is analogous to 
Eq.~(\ref{mnb}) with the pion interchanged from left and to right sides and 
with the necessary permutation 
of quark coordinates ${\b\rho}_2$ and ${\b\rho}^{\prime}_2$ in the operator 
(\ref{e24}). In fact, the final expressions for  matrix elements coincide 
with Eqs.~(\ref{mnp}) and (\ref{ampl}) as far 
as the transition of the valence quark from $N$ to $B$ (with the 
interchange of its number $3 \leftrightarrow 4$) is the same. Of course, 
the pole denominators 
for the diagrams Fig.~\ref{f1}a and Fig.~\ref{f1}b are different.

After averaging over initial spin projections and summing over final spin 
projections, we obtain the squared amplitudes for the virtual subprocess 
$N \to B+\pi$ in the quark model\footnote{The normalization of coupling constants in Eqs.~(\ref{ampl}) and (\ref{const}) for baryons with different spins 
are choose so that their averaged squared amplitudes~(\ref{ms}) have the 
common numerical factor 2 characteristic of the pseudo-scalar $\pi NN$
squared amplitude.}:
\begin{eqnarray}
\overline{|\cal M_s(N\!\to\!\pi\!+\!N)|^2}&=&
2g_{\pi {\scriptscriptstyle NN}}^2 {\bf k}^2
F_{\pi {\scriptscriptstyle NN}}^2({\bf k}^2)
\left[1+\omega_{\pi}({\bf k})/(2M_{\scriptscriptstyle N})\right]^2,
\nonumber\\
\overline{|\cal M_s(N \to \Delta \pi)|^2}&=&
2g_{\pi {\scriptscriptstyle \Delta N}}^2 {\bf k}^2
F_{\pi {\scriptscriptstyle \Delta N}}^2({\bf k}^2)
\left[1+\omega_{\pi}({\bf k})/
(M_{\scriptscriptstyle N}\!+\!M_{\scriptscriptstyle \Delta})\right]^2,
\nonumber\\
\overline{|\cal M_s(N \to N^* \pi)|^2}&=&
2g_{\pi {\scriptscriptstyle N^*N}}^2 \omega_\pi^2({\bf k})
F_{\pi {\scriptscriptstyle N^*N}}^2({\bf k}^2),
\label{e29}
\nonumber\\
\overline{|\cal M_s(N \to N^{**} \pi)|^2}&=&
2g_{\pi {\scriptscriptstyle N^{**}N}}^2 {\bf k}^2
F_{\pi {\scriptscriptstyle N^{**}N}}^2({\bf k}^2)
\left[1+\omega_{\pi}({\bf k})/
(M_{\scriptscriptstyle N}\!+\!M_{\scriptscriptstyle N^{**}})\right]^2.
\label{ms}
\end{eqnarray}
%%%%%%%%%%%%%%%
Using Eqs.~(\ref{ms}), we can write for the 
longitudinal cross-section of the $N \to N+\pi$ channel
\begin{equation}
\frac{d \sigma_{\scriptscriptstyle L}}{dt}=\frac{1}{64 \pi W^2}
\frac{1}{|{\bf q}| |{\bf q}_r^*|}
\frac{2 g_{\pi {\scriptscriptstyle NN}}^2 {\bf k}^2
F_{\pi {\scriptscriptstyle NN}}^2({\bf k}^2)}{(k^2-m_\pi^2)^2}
\left[1\!+\!\frac{\omega_{\pi}({\bf k})}{2M_{\scriptscriptstyle N}}\right]^2
 e^2 F_\pi^2(Q^2)\, \left[(k\!+\!k') \cdot e_{\lambda =0}\right]^2,
\label{e32}
\end{equation}
where $g_{\pi {\scriptscriptstyle NN}}$ and $F_{\pi {\scriptscriptstyle NN}}
=F^{(s)}_{\pi {\scriptscriptstyle NN}}$ are defined by Eqs.~(\ref{fg}) 
and (\ref{ff}) respectively. So, it is possible to determine the only free 
parameter of the model $g_s$ directly from the experiment on the quasi-elastic 
knockout 
of pions and to compare the value $g_{\pi NN}$ of Eq.~(\ref{fg}) obtained by 
this way with the low-energy experimental value $g_{\pi NN}$. The theoretical 
result~(\ref{fg}) is derived 
by comparing expressions for ${\cal M}(N \to N+\pi)$ in the field theory 
[e.g. Eq.~(\ref{e11})] and in the microscopical model [Eq.~(\ref{mnp})]. 
The experimental value $g_{\pi NN}=$ 13.2 was obtained long ago 
from the low-energy $\pi N$ scattering experiment.

%%%%%%%%%%%%%%%%%%%%%%%%%%%%%%%%%%%%%%%%%%%%%%%%%%%%%%%%%%%%%%%%%%%%%%%%%%%%%%%
\subsection{$N \to Y+K$ channels}

When we regard the $N \to Y+K$ channels, in Eq.~(\ref{e24}) the following 
changes should be made: $\tau^{(3)} \to \lambda^{(3)}_{u \to s}$ 
($\lambda^{(3)}_{u \to s}$ is the Gell-Mann matrix corresponding to the 
transition $u \to s$); $g_s \to zg_s$, $m_{\pi} \to m_{\scriptscriptstyle K}$ 
 and $m_q \to m_s$  with 
$m_s\approx{m_{\phi}}/2$ being the mass of strange constituent quark; 
$\omega_\pi ({\bf k}) \to \omega_{\scriptscriptstyle K} ({\bf k})=
\sqrt{{\bf k}^2+m_{\scriptscriptstyle K}^2}$ for kaons on their mass shell
and  $\omega_{\scriptscriptstyle K} ({\bf k})=M_{\scriptscriptstyle N}-
\sqrt{M_{\scriptscriptstyle Y}^2+{\bf k}^2}$ for virtual kaons.. 
Wave functions of the final baryons $Y$ have the form
\begin{equation}
|Y>=|s^3[3]_{\scriptscriptstyle X}L=0>_{\scriptscriptstyle TISM}
|[1^3]_c,([21]_S \circ [21]_F)[3]_{SF}:[1^3]_{CSF}>,
\label{e33}
\end{equation}
where the coordinate (TISM) part coincides with the coordinate part of
nucleon wave function in Eqs.~(\ref{np})-(\ref{norm}). The spin-flavor part 
of Eq.~(\ref{e33}) is similar to the spin-isospin part of neutron wave 
function~(\ref{np}) in which one of d-quarks is replaced with the s-quark.  
The matrix element for the transition  $p \to K+Y$ ($Y=\Lambda,\,\,\Sigma_0$)
is defined in analogy with Eqs.~(\ref{mn1})-(\ref{mnp}) and (\ref{ampl}) 
by the formula:
\begin{eqnarray}
\cal M_s(p\!\to\!K_{\alpha}\!+\!Y)&=&<\!K(q\bar s),\alpha {\bf k}|<\!Y(qqs)|
 H_s|p(3q)\!>=3<\!Y|H^{(3)}_{Kqs(s)}|p\!>,
\label{mnk}
\end{eqnarray}
where $H^{(3)}_{Kqs(s)}$ is derived from $H^{(3)}_{\pi qq(s)}$ by replacing
the pion parameters $m_{\pi}$, $b_{\pi}$ with the kaon ones 
$m_{\scriptscriptstyle K}$, $b_{\scriptscriptstyle K}$ and substituting
$m_q \to m_{s}$. With the standard fractional parentage coefficient 
technique we obtain for the transition matrix element the expression
\begin{eqnarray}
{\cal M}_s({\scriptstyle N}\!\to\!K^{\alpha}\!+\!{\scriptstyle Y})&=&
ig_{\scriptscriptstyle KNY}
F_{\scriptscriptstyle KNY}({\bf k}^2)
\left (\beta {{\cal F}^{({\scriptscriptstyle NY})}_{\alpha}}^{\dagger}+
(1-\beta){{\cal D}^{({\scriptscriptstyle NY})}_{\alpha}}^{\dagger}\right )
\nonumber\\
&\times&{\b\sigma}^{(\scriptscriptstyle N)}\!\cdot\!\!
\left[{\bf k}-\frac{\omega_{\scriptscriptstyle K}(\bf k)}
{M_{\scriptscriptstyle N}\!+\!M_{\scriptscriptstyle Y}}
({\bf P}\!+\!{\bf P}^{\prime})\right],
\label{my}
\end{eqnarray}
which is analogues to Eqs.~(\ref{mnp}) and (\ref{ampl}). Here 
${\cal F}^{({\scriptscriptstyle NY})}_{\alpha}$ and 
${\cal D}^{({\scriptscriptstyle NY})}_{\alpha}$ are matrices of two
non-equivalent (antisymmetric and symmetric) representations of
8-dimensional F-spin ($\alpha=$ 1,2,...,8) for the baryon octet ($N$,
$Y=\Lambda$, $\Sigma$, $\Xi$). In the $^3P_0$ model  the mixing parameter 
has the same value $\beta=$ 2/5 as in the case of SU(6) symmetry 
(see, e.g. Ref.~\cite{fe}), and the $KN\Lambda$ and $KN\Sigma$ coupling 
constants calculated with this parameter,     
\begin{eqnarray}
f_{\scriptscriptstyle KN\Lambda}&=&
-\sqrt{3}\frac{zg_s}{m_s}(2\pi b^2_{\scriptscriptstyle K}
m^2_{\scriptscriptstyle K})^{3/4}
\left[1-\frac{y_{\scriptscriptstyle K}}{3}
\varphi_{\scriptscriptstyle \Lambda}(0)\right]
(1-y_{\scriptscriptstyle K})^{3/2},
\nonumber\\
f_{\scriptscriptstyle KN\Sigma}&=&
\frac{1}{3}\frac{zg_s}{m_s}(2\pi b^2_{\scriptscriptstyle K}
m^2_{\scriptscriptstyle K})^{3/4}
\left[1-\frac{y_{\scriptscriptstyle K}}{3}
\varphi_{\scriptscriptstyle \Sigma}(0)\right]
(1-y_{\scriptscriptstyle K})^{3/2},\nonumber\\
g_{\scriptscriptstyle KN\Lambda}&=&
\frac{M_{\scriptscriptstyle N}\!+\!M_{\scriptscriptstyle \Lambda}}
{m_{\scriptscriptstyle K}}f_{\scriptscriptstyle KN\Lambda},
\quad g_{\scriptscriptstyle KN\Sigma}=
\frac{M_{\scriptscriptstyle N}\!+\!M_{\scriptscriptstyle \Sigma}}
{m_{\scriptscriptstyle K}}f_{\scriptscriptstyle KN\Sigma},
\label{cc}
\end{eqnarray}
have the relative value 
\begin{equation}
f_{\scriptscriptstyle KN\Lambda}/f_{\scriptscriptstyle KN\Sigma}
=-3\sqrt{3}\left[1-\frac{y_{\scriptscriptstyle K}}{3}
\varphi_{\scriptscriptstyle \Lambda}(0)\right]/
\left[1-\frac{y_{\scriptscriptstyle K}}{3}
\varphi_{\scriptscriptstyle \Sigma}(0)\right],
\label{rto}
\end{equation}
which is the consequence of SU(6) symmetry (the factor $-3\sqrt{3}$)~\cite{fe}
violated slightly by the difference of $\Lambda$ and $\Sigma^0$ masses
(the factors in squared brackets). 
Vertex form factors $F_{\scriptscriptstyle KNY}$ have a unified CQM form 
(\ref{ff}) 
\begin{eqnarray}
F_{\scriptscriptstyle KNY}({\bf k}^2)=
\left[1-y_{\scriptscriptstyle K}\varphi_{\scriptscriptstyle Y}(0)/3\right]^{-1}
\left[1-y_{\scriptscriptstyle K}\varphi_{\scriptscriptstyle Y}
({\bf k})/3\right]
exp\left[-\frac{{\bf k}^2b^2}{6}\left(1+y_{\scriptscriptstyle K}/4\right)
\right],
\label{ffs}
\end{eqnarray}
where $y_{\scriptscriptstyle K}=\frac{2}{3}x^2_{\scriptscriptstyle K}
\left[1+\frac{2}{3}x^2_{\scriptscriptstyle K}\right]^{-1}$,
$x_{\scriptscriptstyle K}=b_{\scriptscriptstyle K}/b$, and
$\varphi_{\scriptscriptstyle Y}({\bf k})=3\omega_{\scriptscriptstyle K}
({\bf k})\left[M_{\scriptscriptstyle N}+M_{\scriptscriptstyle Y}+
\omega_{\scriptscriptstyle K}({\bf k})\right]^{-1}$ ($Y=\Lambda$, $\Sigma^0$).
Finally, the averaged square amplitudes for the virtual processes
$p\to K+\Lambda$ and $p\to K+\Sigma_0$ are
\begin{eqnarray}
\overline{|\cal M_s(p\!\to\!K\!+\!\Lambda)|^2}&=&
2g_{\scriptscriptstyle KN\Lambda}^2 
{\bf k}^2F_{\scriptscriptstyle KN\Lambda}^2({\bf k}^2)
\left[1+\omega_{\scriptscriptstyle K}({\bf k})/
(M_{\scriptscriptstyle N}\!+\!M_{\scriptscriptstyle \Lambda})\right]^2,
\nonumber\\
\overline{|\cal M_s(p\!\to\!K\!+\!\Sigma)|^2}&=&
2g_{\scriptscriptstyle KN\Sigma}^2 
{\bf k}^2F_{\scriptscriptstyle KN\Sigma}^2({\bf k}^2)
\left[1+\omega_{\scriptscriptstyle K}({\bf k})/
(M_{\scriptscriptstyle N}\!+\!M_{\scriptscriptstyle \Sigma})\right]^2
\label{msy}
\end{eqnarray}

The momentum distribution of kaons is given by the general expressions 
(\ref{e5})-(\ref{e7}). The 
cross section is given by a formula analogous to Eq.~(\ref{e10}) with 
the electromagnetic form factor of kaon taken from Ref.~\cite{b21}:

\begin{equation}
F_K(Q^2)= \frac{a}{1+\frac{Q^2}{b_1^2}} 
+ \frac{1-a}{(1+\frac{Q^2}{b_2^2})^2},
\label{e43}
\end{equation}
$a=0.398$, $b_1=0.642$ GeV/$c$, $b_2=1.386$ GeV/$c$.

The phenomenological parameter $z$ in Eqs.~(\ref{hs}) and (\ref{cc}) should 
be fitted to the experimental ratio 
$|g_{\scriptscriptstyle KN\Lambda}/g_{\pi{\scriptscriptstyle NN}}|\approx
0.6\div 1.2$, and we shall estimate this value using the data on the 
longitudinal differential cross section for the reaction 
$p(e,e^{\prime}K^{+})\Lambda$ (see next section).
In the scalar $^3P_0$ model this ratio can be extracted from 
Eqs.~(\ref{const}) and (\ref{cc}):
\begin{equation}
f_{\scriptscriptstyle KN\Lambda}/f_{\pi{\scriptscriptstyle NN}}=-z
\frac{3\sqrt{3}}{5}\,\frac{m_q}{m_s}
\left(\frac{b_{\scriptscriptstyle K}m_{\scriptscriptstyle K}}
{b_{\pi}m_{\pi}}\right)^{3/2}
\left[1-\frac{y_{\scriptscriptstyle K}}{3}
\varphi_{\scriptscriptstyle \Lambda}(0)\right]/
\left[1-\frac{y_{\pi}}{3}
\varphi_{\scriptscriptstyle N}(0)\right]
\label{rat}
\end{equation}
Supposing $b_{\scriptscriptstyle K}=b_{\pi}$ and taking into account that
$\left[1-\frac{y_{\scriptscriptstyle K}}{3}
\varphi_{\scriptscriptstyle \Lambda}(0)\right]/\left[1-\frac{y_{\pi}}{3}
\varphi_{\scriptscriptstyle N}(0)\right]\approx$ 1
one can see that the anomalously
small pion mass $m_{\pi}\ll 2m_q$ with respect to the kaon mass
$m_{\scriptscriptstyle K}\approx m_s+m_q$ considerably destroys the 
$SU(3)$-symmetry of meson coupling constants in the scalar $^3P_0$ model.
The value $z\approx$ 0.3 $\div$ 0.5 would be desirable to compensate 
this too large $SU(3)$ violation. (In the next section we estimate this 
value by comparing our predictions with the experimental data).
It should be noted that any microscopic
mechanism of $\bar{q}q$ pairs generation in the QCQ vacuum also leads to 
the $\bar{s}s$ pairs suppression with respect to non-strange pairs owing 
to the large mass of strange quark. One can consider the factor $z$ in the
Hamiltonian (\ref{hs}) as a manifestation of such suppression.

%%%%%%%%%%%%%%%%%%%%%%%%%%%%%%%%%%%%%%%%%%%%%%%%%%%%%%%%%%%%%%%%%%%%%%%%%%
\section{Results and discussion}

\subsection{$N \to N+\pi$ channel}
Various phenomenological versions of momentum distribution (MD) of pions in 
the channel $N\to N+\pi$ are presented in Fig.~\ref{f4}. The solid line 
corresponds to the MD calculated using Eqs.~(\ref{e12}), (\ref{e13}) with 
the experimental value $g_{\pi{\scriptscriptstyle NN}}=$ 13.2. The cut-off
parameter $\Lambda_{\pi}=$ 0.6 GeV/$c$ taken in the calculations reasonably
fits the experimental data (see Figs.~\ref{f5} and \ref{f6}, dashed lines) in 
the area $Q^2\gtrsim$ 1 $(GeV/c)^2$ of $t$-pole bright predominance. 
   
The dashed line in Fig.~\ref{f4} is MD calculated by means of Eq.~(\ref{e16}) 
using Afnan's $\pi N$ potential~\cite{b13}, and the dash-dotted line 
corresponds to MD calculated by means of Eq.~(\ref{e16}) using Lee's 
$\pi N$ potential~\cite{b14}. We see that the latter is rather far from the 
solid line and, consequently, from the experimental data. The calculations 
based on the Afnan's potential fit the experimental data rather well.

One can clearly see from Figs.~\ref{f5} and \ref{f6} that our $t$-pole 
microscopic approach works well\footnote{The surprising thing is that it 
concerns the larger kinematic region ${\bf k}^2\lesssim$ 0.5 $GeV^2/c^2$ 
than the region of quasi-elastic knockout kinematics 
${\bf k}^2\lesssim$ 0.3 $GeV^2/c^2$.} at $Q^2\gtrsim$ 1 $(GeV/c)^2$: both the
absolute value of cross section $d\sigma_{\scriptscriptstyle L}/dt$ and the
shape of its dependence on $t=-{\bf k}^2$ (i.e. the shape of 
$|\Psi_p^{n\pi^+}({\bf k}^2)|^2$) are well reproduced by the microscopic
theory normalized to the low-energy pion-nucleon data (solid line). At the 
same time at $Q^2\lesssim$ 1 $(GeV/c)^2$ the pole approximation is not 
efficient, what is in a good agreement with our previous 
estimations~\cite{b5} (the disturbing influence of the competing $s$-pole 
diagram and of the $t$-pole diagram with a virtual $\rho$-meson becomes to 
be essential here~\cite{b5}).

Further, our discussed results show (see Fig.~\ref{f7}a) that both the shape 
of phenomenological monopole form factor (\ref{e13}) and the empirical value
0.6$\div$0.7 GeV/c of the cut-off parameter $\Lambda_{\pi}$ (it appears 
close to that by Phandaripande et al.~\cite{phan} connected to a
phenomenological analysis of independent problems) find its microscopic
foundation. In particular, it means that the rms radius of the nucleon is 
reproduced correctly by these soft cut-off parameters. 

So, our predictions for $N\to B+\pi$ channels ($B=\Delta, N_{1/2^-}(1535),
N_{1/2^+}(1440)$) seem to be useful estimates for future exclusive 
experiments.

%%%%%%%%%%%%%%%%%%%%%%%%%%%%%%%%%%%%%%%%%%%%%%%%%%%%%%%%%%%%%%%%%%%%%%%%%
\subsection{$N \to B+\pi$ channels in the microscopic model}

Form factors~(\ref{nff}) calculated within the quark $^3P_0$ model for all
considered channels $N\to B+\pi$ are presented in Fig.~\ref{f7}(a). 
For comparison, we also show here for the 
$N\to N+\pi$ channel the monopole form factor~(\ref{e13}) for 
$\Lambda_{\pi}=$ 0.6 and 0.7 GeV/$c$ (dotted lines), which was discussed 
above. The $N \to \Delta +\pi$ channel is characterized 
by the same form factor as the $N \to N +\pi$ channel. However, the MD will 
be different, first, because of the difference in the pole denominators in 
Eq.~(\ref{e5}) (for $B=\Delta$, $N^*$, $N^{**}$ these denominators are bigger 
than for $B=N$ and are almost ${\bf k}^2$-independent). Second, the vertex 
constants for the $p\to\pi+n$ and $p\to\pi+\Delta$ channels are also 
different. For comparison the pion momentum distributions for these channels 
are shown in Fig.~\ref{f8}.

Of particular interest is the dependence of $NB\pi$ vertex functions on the
pion radius $b_{\pi}$.
The general feature is that all form factors for transitions without radial or
orbital excitation of the quark wave function (e.g., $N\to N$, $N\to \Delta$, 
$N\to \Lambda$) are practically independent from the parameter 
$x_\pi=b_\pi/b$ ($x_{\scriptscriptstyle K}=b_{\scriptscriptstyle K}/b$).
However, the vertex constants
depend strongly on $x_\pi$. The ratio $R=g_{\pi BN}^2/g_{\pi NN}^2$
calculated within the microscopical model considered for the channels
$B=N^*$, $N^{**}$ is presented in Fig.~\ref{f7}(b) as a function of $x_\pi$.
The measurements of the $p(e,e' \pi)B$
cross-sections will provide us, in particular, with the relative values
$g_{\pi BN}^2/g_{\pi NN}^2$.
The comparison of the predicted  ratios to the experiment should be essential
in the verification of this model including independent obtaining of
the $x_\pi$ value  and its comparison with that which follows from the
shape of the pion electromagnetic form factor [in this work we use only
the last experimental data~\cite{b12b} on the pion off-shell charge form 
factor $F_{\pi}(Q^2)$].

For the channel $B=\Delta$ the ratio $R$ does not depend on $x_\pi$,
$R=(3\sqrt{2}(M_{\scriptscriptstyle \Delta}+M_{\scriptscriptstyle N})
/5M_{\scriptscriptstyle N})^2=$ 3.85.
We have obtained the following values of $NB\pi$ coupling constants  for 
the point-like pion ($x_\pi=0$): $g_{\pi \Delta N}^2/4 \pi=53.4$,
$g_{\pi N^*N}^2/4 \pi=11.8$ and $g_{\pi N^{**} N}^2/4 \pi=$ 2.1, while at
$x_{\pi}=0.5$ the $\pi NN^{**}$ coupling constant becomes larger:
$g_{\pi N^{**} N}^2/4 \pi=$ 3.09 (see Fig.~\ref{f7}(b)).

The form factors for transitions with orbital or radial excitation (i.e. for 
the $B=N^*$ and $N^{**}$ channels) depend on the $x_{\pi}$. Assuming 
$x_{\pi}=$ 0.5 we have obtained that these form factors  change the 
sign at $|{\bf k}^2| \simeq$ 1.22 and 0.72 (GeV/$c$)$^2$. It could be 
considered as a manifestation of the excited quark configurations $s^2p$ and 
$sp^2$ in baryons $N^*$ and $N^{**}$ respectively. 
This is a directly observable effect.

%%%%%%%%%%%%%%%%%%%%%%%%%%%%%%%%%%%%%%%%%%%%%%%%%%%%%%%%%%%%%%%%%%%%%%%%%%%%
\subsection{$N \to Y+K$ channels}

Our calculations of the MDs of kaons in the channels $p \to 
\Lambda +K$ and $p \to \Sigma +K$ within the $^3P_0$ model are shown 
in Fig.~\ref{f9}. Predicted spectroscopic factors are 
$S^{K \Lambda}_p=0.152$ and $S^{K \Sigma}_p=0.006$ 
(for comparison $S^{\pi N}_N=0.25$). The corresponding cross sections are 
presented in Fig.~\ref{f10}.

Unfortunately, the longitudinal $d\sigma_{\scriptscriptstyle L}/dt$ 
and transverse $d\sigma_{\scriptscriptstyle T}/dt$
differential cross sections of the channel $N \to Y+K$ are not separated in 
the available data, and we cannot extract the MD of kaons from the experiment 
as we did it for pions. Ref.~\cite{b12a} gives us only the ratio of these 
cross sections $R=(d\sigma_{\scriptscriptstyle L}/dt)/
(d\sigma_{\scriptscriptstyle T}/dt)\approx$ 1
in the narrow range $t=0.33-0.38$ $GeV^2/c^2$ for $\Lambda+K^+$ channel and 
$t=0.25-0.3$ $GeV^2/c^2$ for $\Sigma^{0}+K^+$ channel. Recently~\cite{b24a} 
the ratio $R^{\prime}=(d\sigma_{\scriptscriptstyle L}/d\Omega)/
(d\sigma_{\scriptscriptstyle T}/d\Omega)\approx$ 0.5  was measured in the
$p(e,e^{\prime}K^{+})\Lambda$ reaction for zero angle 
$\theta_{\scriptscriptstyle K}$ of kaon emission 
with respect to the photon momentum ${\bf q}^{cm}$ in the area 0.5 
$< Q^2 <$ 2. $(GeV/c)^2$. 

Having in mind the large spread of experimental data, we consider here two
variants, $R=$ 0.5 and $R=$ 1. Using the old DESY data~\cite{b12} on 
$d\sigma/dt=\varepsilon d\sigma_L/dt +d\sigma_T/dt$ for $Q^2=$ 0.7 and 1.35
$(GeV/c)^2$ at $\varepsilon=$ 0.85 and 0.82 (W = 1.9 $\div$ 2.5 GeV)
for both $p\to \Lambda+K^+$ and $p\to \Sigma^0+K^+$ channels
we have estimated the longitudinal cross section as
$d\sigma_{L}/dt=\frac{R}{R+\varepsilon}\,d\sigma/dt$. With these data for
$\Lambda+K^+$ channel (see Fig.~\ref{f10}(a,b)) we have estimated the value 
of the parameter $z$ in Eq.~(\ref{rat}), $z\approx$ 0.4$\div$0.5. In 
Figs.~\ref{f10}(a-d) and \ref{f11}(a,b) the predictions of the scalar 
$^3P_0$ model for longitudinal cross sections of quasi-elastic kaon knockout 
from the proton are shown (solid line) for both channels $\Lambda+K$ and
$\Sigma^0+K$ with using the value $z=$ 0.5.
 
The $Q^2$-dependence of the differential cross sections $d\sigma_L/d\Omega$
for $\theta_{\scriptscriptstyle K}=0^o$ is shown on Fig.\ref{f11} (solid 
line). For this dependence there exists separated experimental data 
for the longitudinal and transverse differential cross sections~\cite{b24a}, 
but this dependence does not give us the shape of the MDs. 
For example, at $Q^2\gtrsim$ 1 $(GeV/c)^2$ these zero-angle data correspond 
to too large virtual-pion momenta ${\bf k}^2\gtrsim$ 0.5 $(GeV/c)^2$ to be 
described in terms of merely kaon t-pole diagrams. 

Note, that the theoretical calculations within the $^3P_0$ model do not 
reproduce the ratio of the longitudinal cross sections for the 
$N \to \Lambda +K$ and $N \to \Sigma +K$ channels. The theoretical result is
$d\sigma_L/dt(N \to \Lambda +K): d\sigma_L/dt(N \to \Sigma +K) \approx$
$g_{K \Lambda N}^2/g_{K \Sigma N}^2 =27$ (in fact, the ratio of the cross 
sections is more than 27 due to the difference in mass between $\Lambda$ and 
$\Sigma$ baryons; it is about 40). The experimental ratio is much smaller.
This may be an indication of the fact that another mechanism  contributes 
considerably to the cross section. For example, an enhancement of the
$K\Sigma^0$ amplitude as it was pointed out in Ref.~\cite{b24a} may originate 
from s-channel $\Delta^*$ resonances which are forbidden by isospin in the 
$K\Lambda$ system.

Recall that the ratio $g_{K \Lambda N}^2/g_{K \Sigma N}^2 =27$ within the 
scalar $^3P_0$ model is a consequence of Eqs.~(\ref{cc}) and (\ref{rto}), 
which, in turn, is a result of $SU(6)$ spin-flavor symmetry of initial 
interaction Hamiltonian~(\ref{hs}). To obtain absolute values of these 
constants we must normalize the only free parameter of the model, $g_s$, 
to a known vertex constant $g_{\pi NN}$, and it gives 
$g_{\scriptscriptstyle K \Lambda N}/\sqrt{4 \pi} =$ -5.1 ,
$g_{\scriptscriptstyle K \Sigma N}/\sqrt{4\pi} =$ 1.0, if $z=$ 1, but in 
a more realistic case, when $z=$ 0.5, it gives 
$g_{\scriptscriptstyle K \Lambda N}/\sqrt{4 \pi} =-2.53$ ,
$g_{\scriptscriptstyle K \Sigma N}/\sqrt{4\pi} =0.51$.
For comparison, Refs.~\cite{b21,b22} give us the following values:
$g_{\scriptscriptstyle K \Lambda N}/\sqrt{4 \pi} =-3.16$, 
$g_{\scriptscriptstyle K \Sigma N}/\sqrt{4 \pi} =0.91$ 
\cite{b21} and $g_{\scriptscriptstyle K \Lambda N}/\sqrt{4 \pi} =-4.17$, 
$g_{\scriptscriptstyle K \Sigma N}/\sqrt{4 \pi} =1.18$~\cite{b22}. 

To refine these values and to clarify the origin of enhancement of 
the $p\to \Sigma+K$ cross section the measurement of kaonic MDs in the
channels $p\to \Lambda+K$ and $p\to \Sigma+K$ (both their normalizations and
shapes) should be very important.

%%%%%%%%%%%%%%%%%%%%%%%%%%%%%%%%%%%%%%%

\subsection{Some perspectives}

Basing on the microscopical $^3P_0$ model we have predicted the pionic MDs,
both the shape of $k$-dependence and normalization, i.e. spectroscopic
factor, in the channels of virtual decay $N \to B+\pi$, $B=N$, 
$\Delta$, $N^*$, $N^{**}$, and kaonic MDs in the channels $N \to Y+K$, 
$Y=\Lambda$, $\Sigma$.
The microscopic consideration can be easily extended to higher excited states
of the baryon-spectator $B$ 
using the more advanced relativistic version of the $^3P_0$ model. These MDs
can be efficiently verified in the quasi-elastic knockout experiments 
$p(e,e' \pi)B$ and $p(e,e'K)Y$ at the electron beam energy of a few GeV. 
Such broad verification should be an important ground for further improvement 
of microscopical quark models, which provide us with the valuable universal 
basis for consideration of various meson clouds. Our nearest intention here 
consists in consideration of the transverse cross section $d\sigma_T/dt$ of 
the process $p(e,e' \pi)B$. It opens a way to the investigation of the 
$\rho$-meson MD $\overline{|\Psi_p^{n \rho}({\bf k})|^2}$
via very predominating (at $Q^2=2-4$ GeV$^2$/$c^2$) virtual 
subprocess $\rho^* +\gamma_T^* \to \pi$ ($*$ means a virtual particle).
Analogously, we plan to calculate microscopically the MDs of vector strange
mesons which can be extracted from the transverse cross sections 
$d\sigma_{\scriptscriptstyle T}/dt$ of $p(e,e^{\prime})Y$ quasi-elastic
knockout processes.

Further, as a next step, an amplitude of a non-diagonal subprocess 
$\pi^*(L) +\gamma^* \to \pi (L=0)$ with rearrangement of the nonzero 
internal orbital momentum $L$ of the virtual pion $\pi^*$ to $L=0$ can be 
extracted from the properly organized experiment, if we take into account 
our experience in the physics of clusters in the nucleus. Namely, the 
theory of quasi-elastic knockout of clusters from nuclei by intermediate 
energy protons \cite{b4,b16} shows that the amplitudes of deexcitation of 
virtually excited clusters [say, $\alpha^*(L \ne 0)+p \to \alpha 
(L=0)+p'$] play very important role here. These amplitudes can be 
revealed by very large anisotropies of differential cross sections in 
regards to the orientation of the recoil momentum ${\bf k}$ of the 
final nucleus-spectator $A-4$ with respect to the bombarding beam 
direction ($\theta$-anisotropy) and to proton scattering plane 
($\phi$-anisotropy, i.e. Treiman-Yang anisotropy~\cite{b4,b16}). For 
electrons with only single electron-quark collisions, the effect will be 
probably not so big, but quite observable~\cite{b24}.

It seems reasonable to join the above microscopic approach with virtually 
excited mesons in the nucleon and the experience of the phenomenological 
theory of Regge poles as applied to the analysis of electroproduction of 
pions and kaons~\cite{b25}. Such synthesis can illuminate microscopical 
ground of the Regge theory.

The present paper, where we use a "naive" non-relativistic model of 
scalar $^3P_0$ fluctuation, aims to outline preliminarily a new horizon in 
both experimental and theoretical intermediate energy research of 
nonpertubative QCD as a basis for detailed explanations and predictions 
of various meson-baryon degrees of freedom.

\bigskip
\centerline{\bf Acknowledgments}
We gratefully acknowledge very useful discussions with Prof. 
A. Faessler and Dr. V. Lyubovitskij.
 This work was partially supported by the Russian Foundation for Basic 
Research (grant N 03-02-17394) and the Deutsche Forshungsgemainschaft 
(grant Fa67/20-1).

\newpage

\begin{figure}[hp]\centering
\epsfig{file=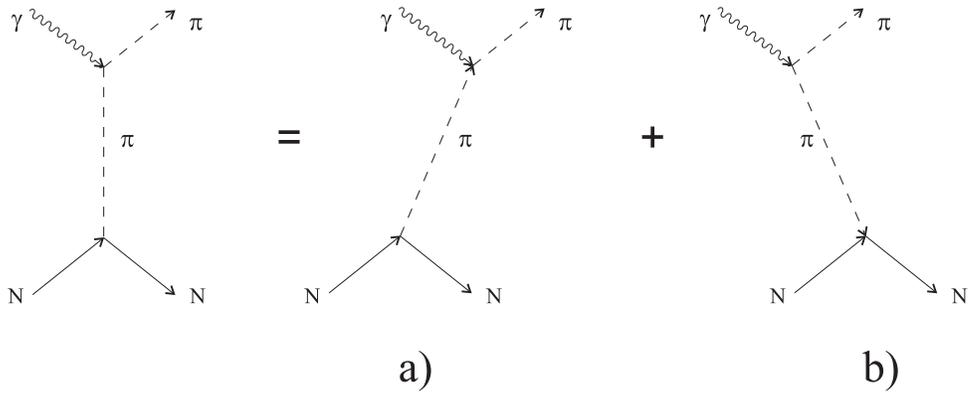,width=0.8\textwidth}
\caption{Feynman $t$-pole diagram for the pion production off the nucleon: 
a) the quasi-elastic knockout of pion; b) $z$-diagram of $\pi^+\pi^-$ 
production.}
\label{f1}
\end{figure}

\centerline{\large \bf Figure captions}
\medskip
{\bf Fig.\ref{f1}.} Feynman $t$-pole diagram for the pion production off 
the nucleon: a) the quasi-elastic knockout of pion; b) $z$-diagram of 
$\pi^+\pi^-$ production.

\begin{figure}[hp]\centering
\epsfig{file=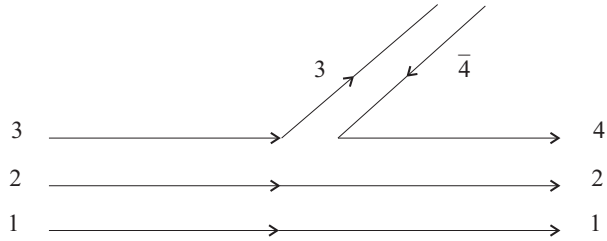,width=0.5\textwidth}
\caption{Quark microscopic picture of pion production inside the
nucleon.}
\label{f2}
\end{figure}

{\bf Fig.\ref{f2}.} Quark microscopic picture of pion production inside the
nucleon.

\begin{figure}[hp]\centering
\epsfig{file=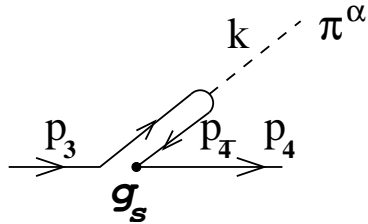,width=0.3\textwidth}
\caption{The effective quark-pion vertex of the scalar $^3P_0$ model.}
\label{f3}
\end{figure}

{\bf Fig.\ref{f3}.} The effective quark-pion vertex of the scalar $^3P_0$ 
model.

\begin{figure}[hp]\centering
\epsfig{file=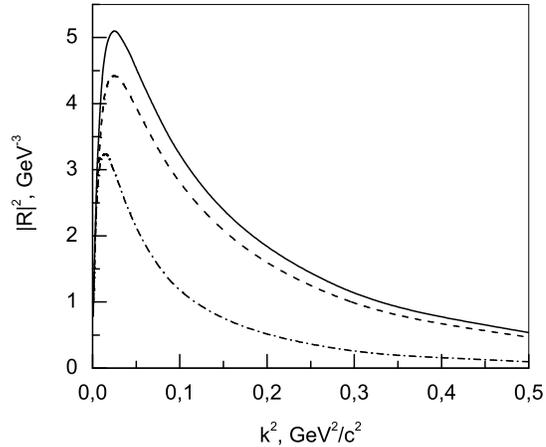,width=0.5\textwidth}
\caption{\label{f4} Momentum distribution of pions in the nucleon. The 
``radial'' part $|R_p^{n \pi^+}({\bf k}^2)|^2$, $(GeV/c)^{-3}$, 
versus ${\bf k}^2$: the solid line for the wave function 
(\ref{e12}); the dashed line for the function (\ref{e16}) calculated on the 
basis of Ref.~\cite{b13} model; the dash-dotted line for the function 
(\ref{e16}) calculated on the basis of Ref.~\cite{b14} model.}
\end{figure}

{\bf Fig.\ref{f4}.} Momentum distribution of pions in the nucleon. The 
``radial'' part $|R_p^{n \pi^+}({\bf k}^2)|^2$, $(GeV/c)^{-3}$,  
versus ${\bf k}^2$: the solid line for the wave function (\ref{e12}); 
the dashed line for the function (\ref{e16}) calculated on the 
basis of Ref.~\cite{b13} model; the dash-dotted line for the function 
(\ref{e16}) calculated on the basis of Ref.~\cite{b14} model. 

{\bf Fig.5.} Longitudinal cross section $d \sigma_L/dt$ for the 
$p(e,e'\pi^+)n$ process calculated using Eqs.~(\ref{e10}) and (\ref{e12}) 
with two different values of $\Lambda_{\pi}=$ 0.6 GeV/$c$ (dashed lines) and 
1.2 GeV/$c$ (dash-dotted lines). Solid lines: predictions of the scalar 
$^3P_0$ model ($b_{\pi}=$ 0.3, $b=$ 0.6 fm). 
Experimental data for $Q^2=$ 0.7 (W=2.19 GeV) from Ref.~\cite{b12} (a) and for 
3.3 $GeV^2/c^2$ (W=2.65 GeV) from Ref.~\cite{b12a} (b).

\begin{figure}
\begin{center}
\mbox{\subfigure[]
{\epsfig{figure=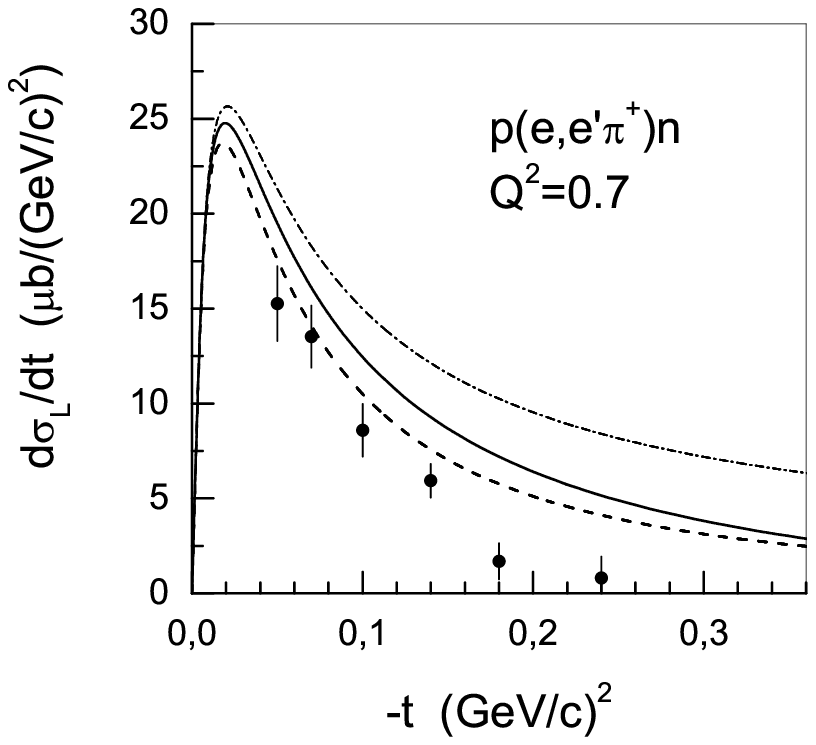,width=0.41\textwidth,clip}}\quad
\subfigure[]{\epsfig{figure=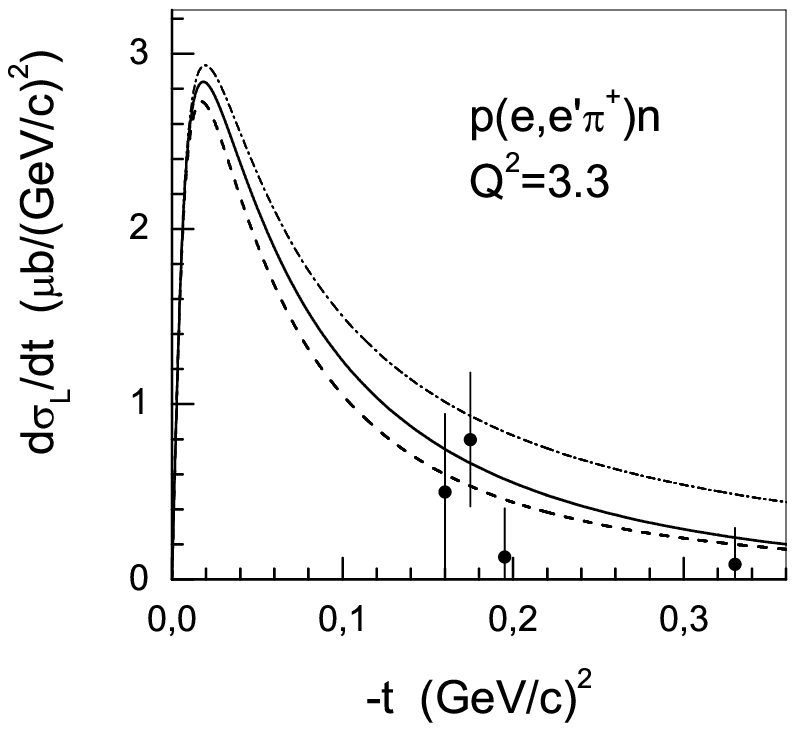,width=0.4\textwidth,clip}}
}
\caption{Longitudinal cross section $d \sigma_L/dt$ for the $p(e,e'\pi^+)n$
process calculated using Eqs.~(\ref{e10}) and (\ref{e12}) with two different 
values of $\Lambda_{\pi}=$ 0.6 GeV/$c$ (dashed lines) and 
1.2 GeV/$c$ (dash-dotted lines). Solid lines: predictions of the scalar 
$^3P_0$ model ($b_{\pi}=$ 0.3, $b=$ 0.6 fm). 
Experimental data for $Q^2=$ 0.7 (W=2.19 GeV) from Ref.~\cite{b12} (a) and for 
3.3 $GeV^2/c^2$ (W=2.65 GeV) from Ref.~\cite{b12a} (b).}
\label{f5}
\end{center}
\end{figure}

\begin{figure}
\begin{center}
\mbox{\subfigure[]
{\epsfig{figure=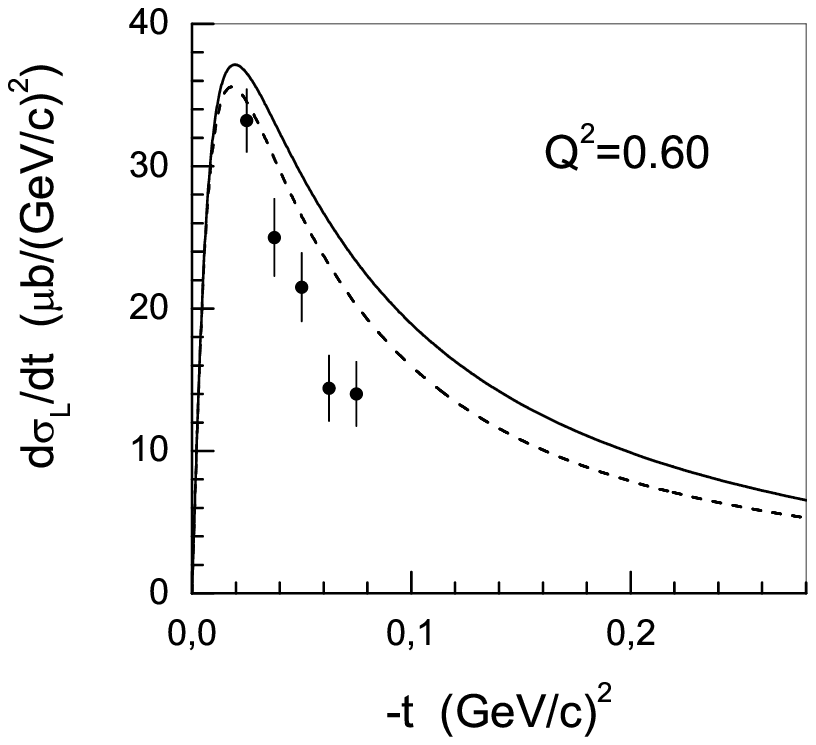,width=0.41\textwidth,clip}}\quad
\subfigure[]{\epsfig{figure=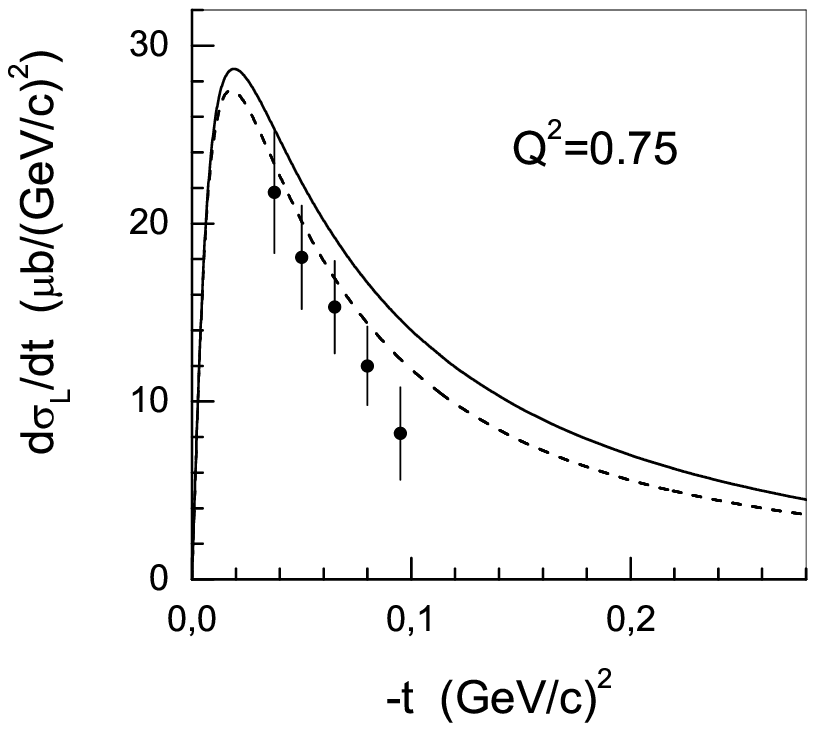,width=0.4\textwidth,clip}}}
\mbox{\subfigure[]
{\epsfig{figure=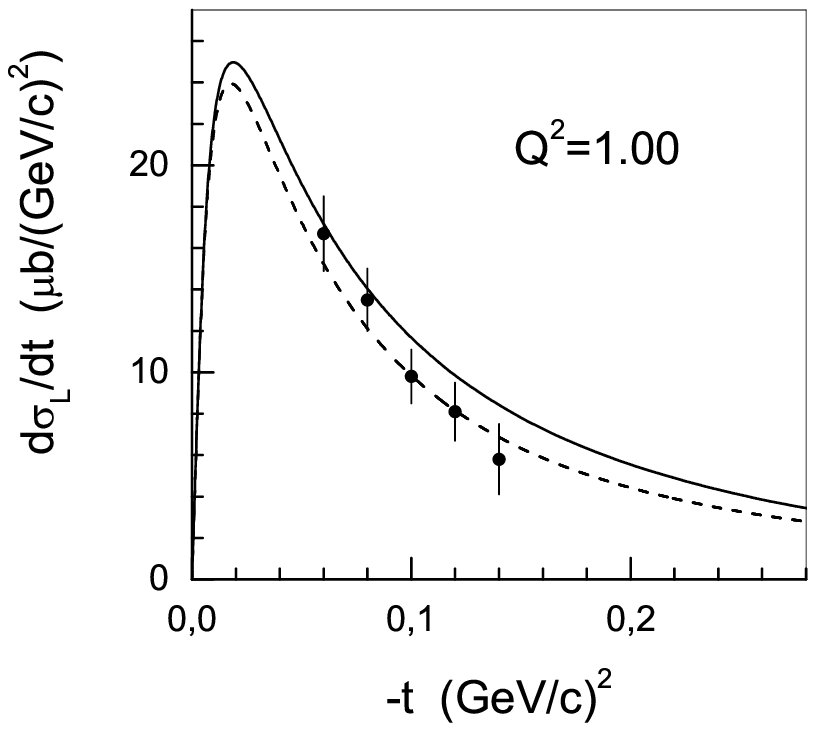,width=0.4\textwidth,clip}}\quad
\subfigure[]{\epsfig{figure=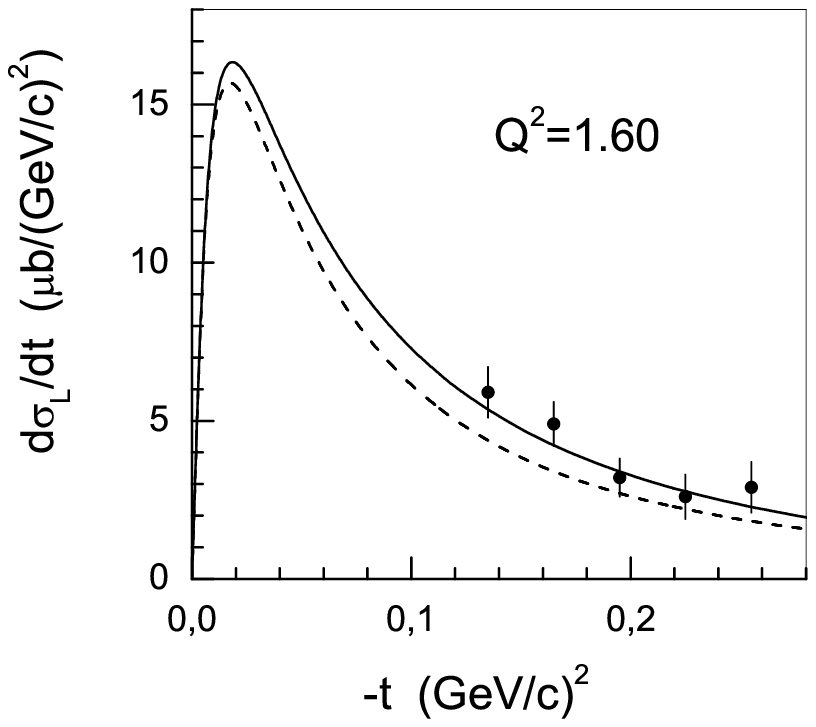,width=0.4\textwidth,clip}}
}
\caption{Longitudinal cross section $d \sigma_L/dt$ for the $p(e,e'\pi^+)n$
process. The same notations as in Fig.~\ref{f5}. Experimental data
from Ref.~\cite{b12b}. $Q^2=$ 0.6 (a), 0.75 (b), 1.0 (c), and 1.6 (d) 
$GeV^2/c^2$, W=1.95 GeV.}
\label{f6}
\end{center}
\end{figure}

{\bf Fig.\ref{f6}.} Longitudinal cross section $d \sigma_L/dt$ for the 
$p(e,e'\pi^+)n$ process. The same notations as in Fig.~\ref{f5}. Experimental 
data from Ref.~\cite{b12b}. $Q^2=$ 0.6 (a), 0.75 (b), 1.0 (c), and 1.6 (d) 
$GeV^2/c^2$, W=1.95 GeV.

\begin{figure}
\begin{center}
\mbox{\subfigure[]
{\epsfig{figure=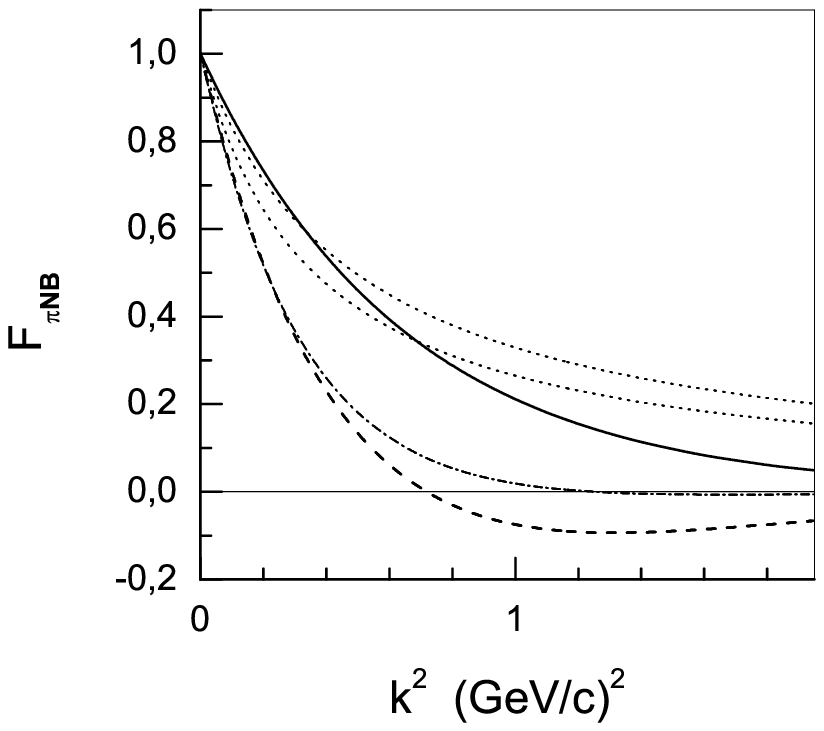,width=0.38\textwidth,clip}}\quad
\subfigure[]{\epsfig{figure=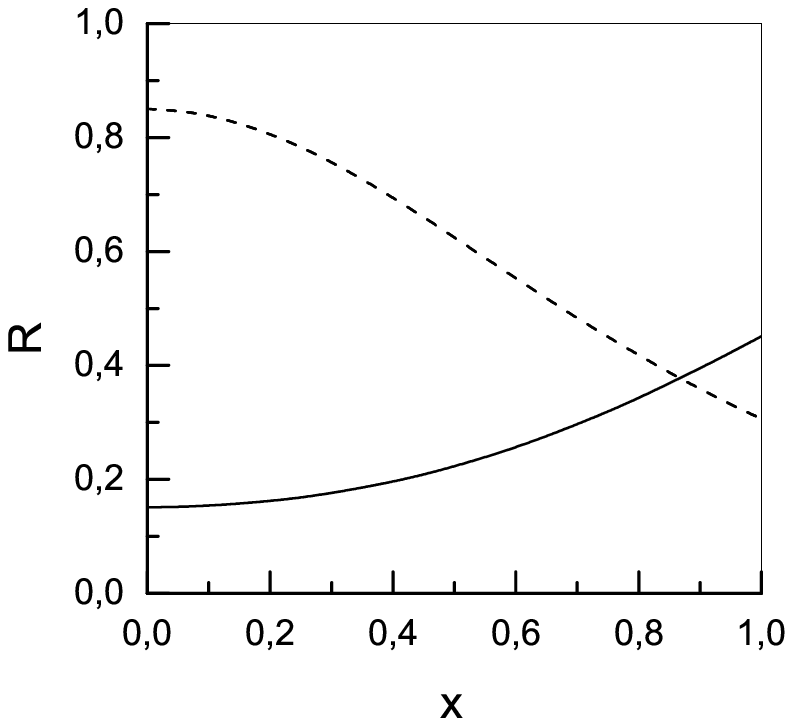,width=0.35\textwidth,clip}}
}
\caption{(a) Strong form factors in the scalar $^3P_0$ model 
($b_{\pi}=$ 0.3 fm, $b=$ 0.6 fm): $F_{\pi NN}$ (solid line), $F_{\pi NN^{**}}$ 
(dashed line), $F_{\pi NN^*}$ (dash-dotted line). Dotted lines correspond 
to the form factor parametrized by Eq.~(\ref{e13}) with 
$\Lambda_{\pi}=$ 0.7 $GeV^2/c^2$ (upper line) and 0.6 $GeV^2/c^2$ (lower line).
(b) The scalar $^3P_0$ model. The ratio of strong coupling constants R 
$=g^2_{\pi N B}/g^2_{\pi NN}$ for $B=N^{**}\equiv N_{1/2^+}(1440)$  
(solid line) and for $B=N^*\equiv N_{1/2^-}(1535)$ (dashed line); 
$x=b_{\pi}/b$.}
\label{f7}
\end{center}
\end{figure}

{\bf Fig.\ref{f7}.} (a) Strong form factors in the scalar $^3P_0$ model 
($b_{\pi}=$ 0.3 fm, $b=$ 0.6 fm): $F_{\pi NN}$ (solid line), $F_{\pi NN^{**}}$ 
(dashed line), $F_{\pi NN^*}$ (dash-dotted line). Dotted lines correspond 
to the form factor parametrized by Eq.~(\ref{e13}) with 
$\Lambda_{\pi}=$ 0.7 $GeV^2/c^2$ (upper line) and 0.6 $GeV^2/c^2$ (lower line).
(b) The scalar $^3P_0$ model. The ratio of strong coupling constants R 
$=g^2_{\pi N B}/g^2_{\pi NN}$ for $B=N^{**}\equiv N_{1/2^+}(1440)$  
(solid line) and for $B=N^*\equiv N_{1/2^-}(1535)$ (dashed line); $x=b_{\pi}/b$

\begin{figure}[hp]\centering
\epsfig{file=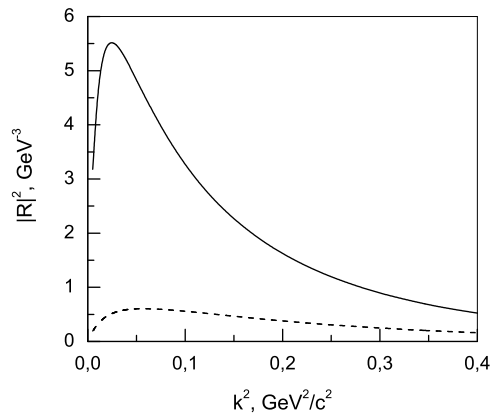,width=0.45\textwidth}
\caption{Momentum distribution of pions in the proton;
solid line: $p \to \pi+n$ channel, 
dashed line: $p \to \pi+\Delta$ channel. The scalar $^3P_0$ model.}
\label{f8}
\end{figure}

\begin{figure}[hp]\centering
\epsfig{file=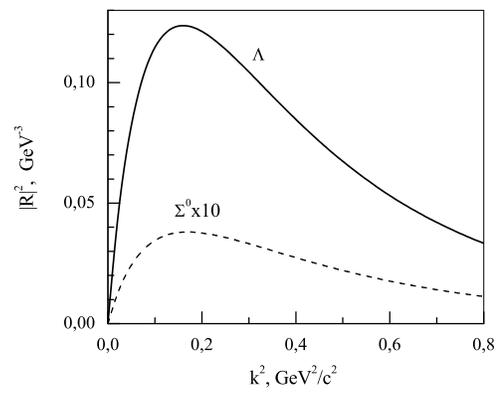,width=0.45\textwidth}
\caption{ Momentum distribution of kaons in the proton; solid line: 
the $p \to K+\Lambda$ channel, dashed line: the $p \to K+\Sigma$ channel 
(times 10). The scalar $^3P_0$ model.}
\label{f9}
\end{figure}

\begin{figure}
\begin{center}
\mbox{\subfigure[]
{\epsfig{figure=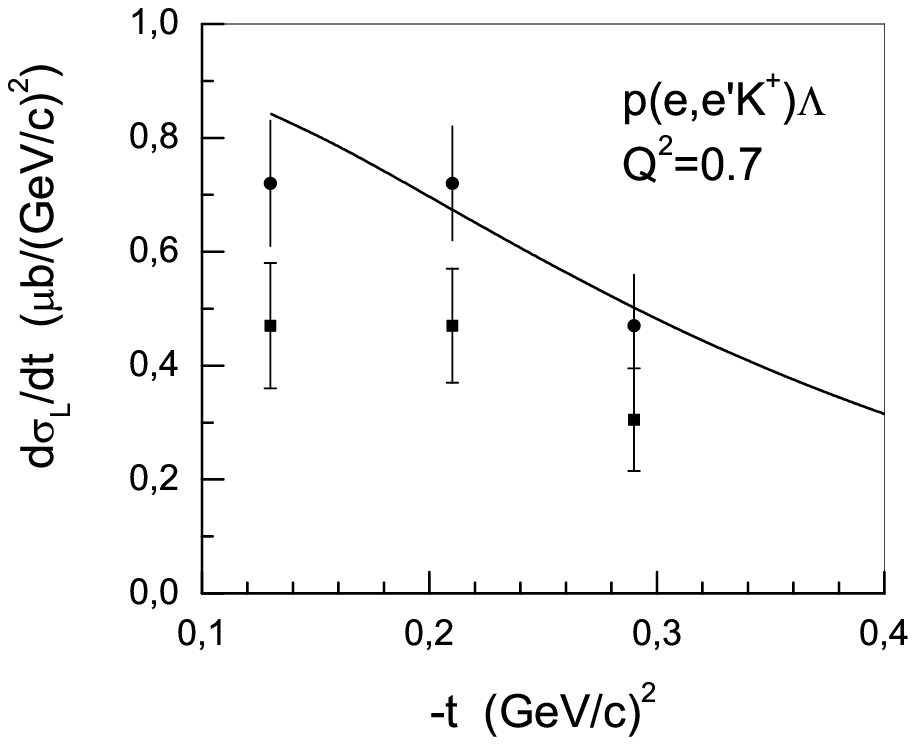,width=0.405\textwidth,clip}}\quad
\subfigure[]{\epsfig{figure=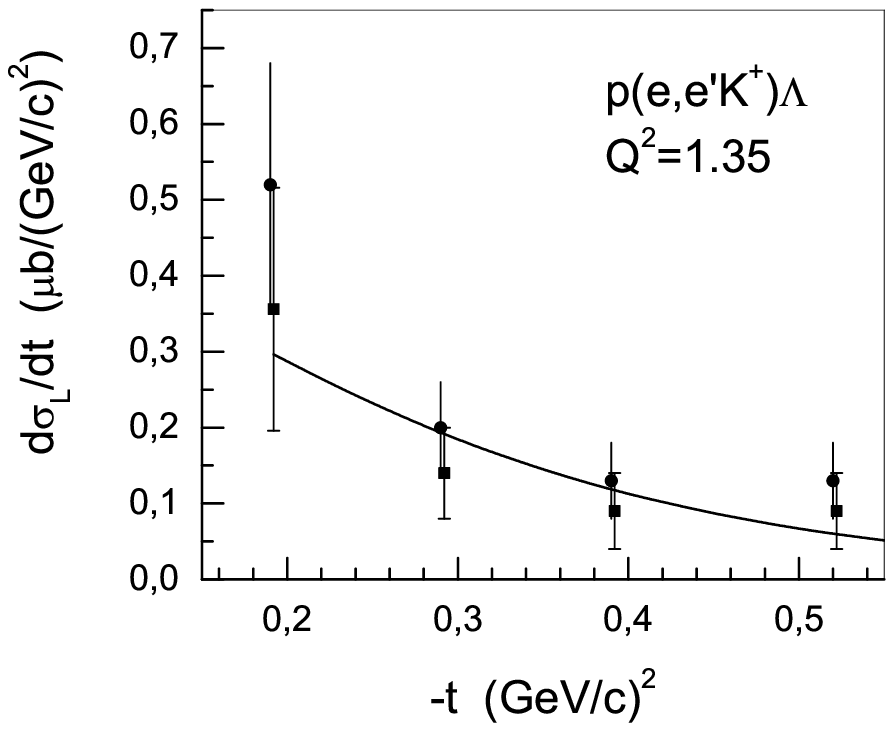,width=0.4\textwidth,clip}}}
\mbox{\subfigure[]
{\epsfig{figure=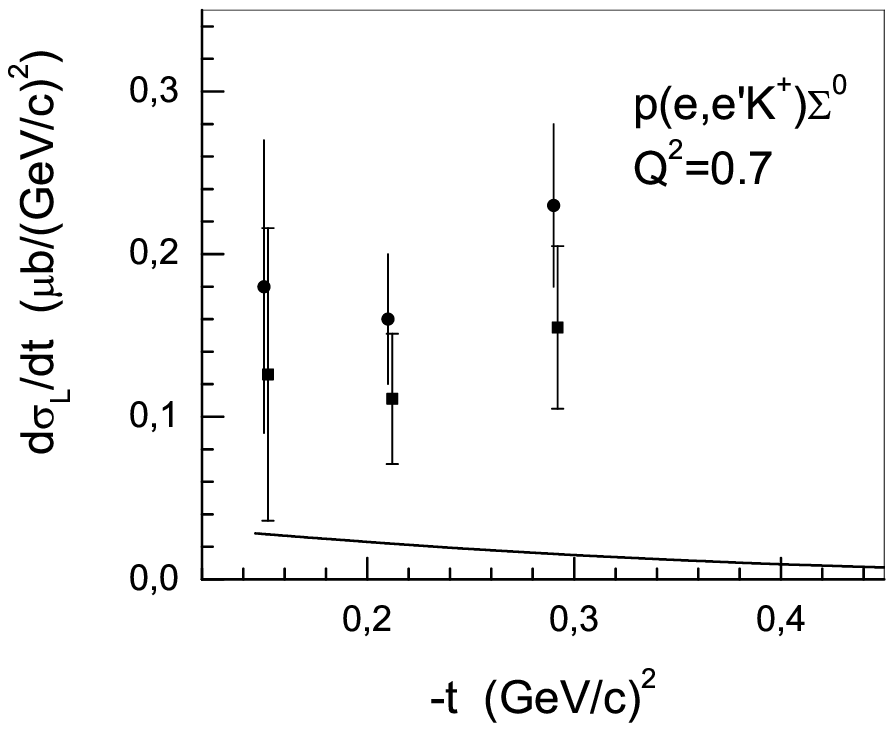,width=0.415\textwidth,clip}}\quad
\subfigure[]{\epsfig{figure=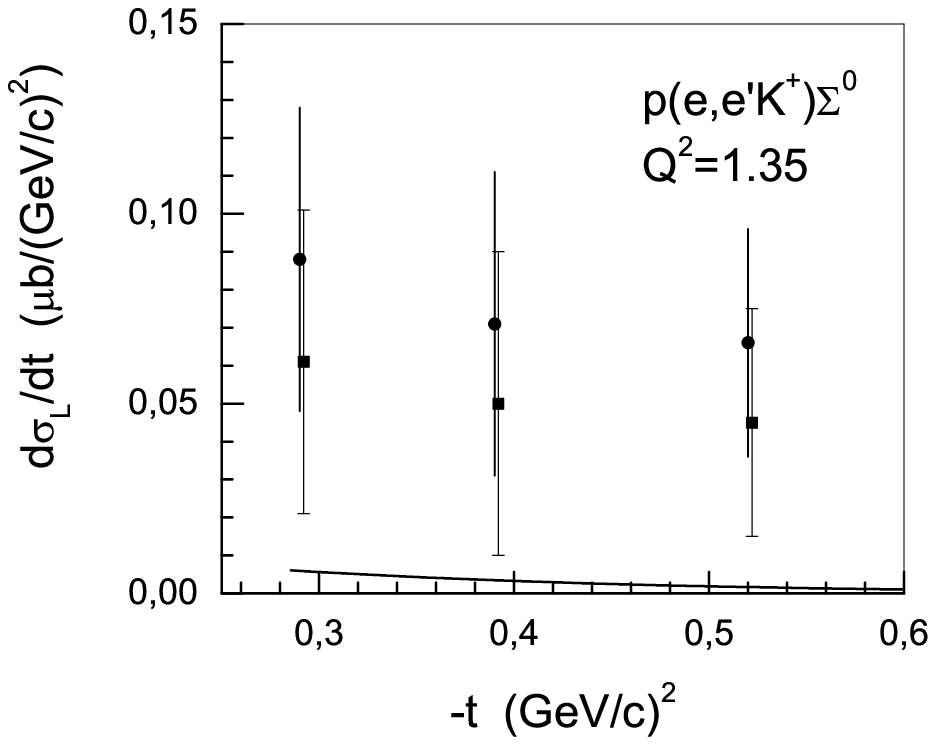,width=0.44\textwidth,clip}}
}
\caption{Kaon quasi-elastic knockout cross sections 
$p(e,e'K^+)\Lambda$ (a,b) and $p(e,e'K^+)\Sigma^0$ (c,d).
Solid lines: the scalar $^3P_0$ model ($b_{\pi}=$ 0.3 fm, $b=$ 0.6 fm, 
$z=$ 0.5). Data from P.Brauel et al.~\cite{b12} (modified): circles for 
$R\equiv\sigma_L/\sigma_T=$ 1. and squares for $R=$ 0.5;
$Q^2=$ 0.7 $GeV^2/c^2$ (a,c), $Q^2=$ 1.35 $GeV^2/c^2$ (b,d); 
$W=$ 1.9$\div$2.5 GeV.}
\label{f10}
\end{center}
\end{figure}

\begin{figure}
\begin{center}
\mbox{\subfigure[]
{\epsfig{figure=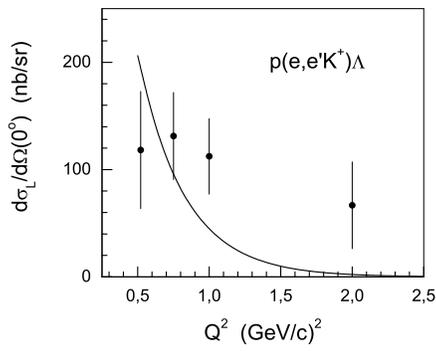,width=0.4\textwidth,clip}}\quad
\subfigure[]{\epsfig{figure=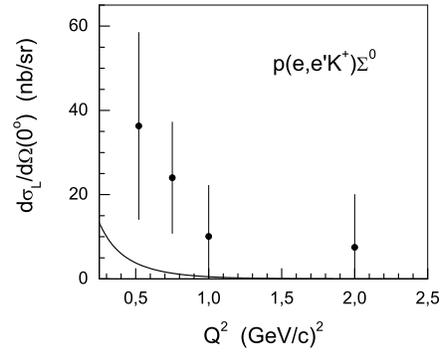,width=0.4\textwidth,clip}}
}
\caption{$Q^2$-dependence of the longitudinal kaon quasi-elastic knockout 
differential cross section $d\sigma_L/d\Omega$ at zero angle to the photon 
direction: (a) $p(e,e'K^+)\Lambda$, (b) $p(e,e'K^+)\Sigma^0$.
Solid lines: the scalar $^3P_0$ model ($b_{\pi}=$ 0.3, $b=$ 0.6 fm, $z=$ 
0.5). Data from Ref.~\cite{b24a}.}
\label{f11}
\end{center}
\end{figure}

{\bf Fig.\ref{f8}.} Momentum distribution of pions in the proton;
solid line: $p \to \pi+n$ channel, 
dashed line: $p \to \pi+\Delta$ channel. The scalar $^3P_0$ model.

{\bf Fig.9.} Momentum distribution of kaons in the proton; solid line: 
the $p \to K+\Lambda$ channel, dashed line: the $p \to K+\Sigma$ channel 
(times 10). The scalar $^3P_0$ model.

{\bf Fig.10.} Kaon quasi-elastic knockout cross sections 
$p(e,e'K^+)\Lambda$ (a,b) and $p(e,e'K^+)\Sigma^0$ (c,d).
Solid lines: the scalar $^3P_0$ model ($b_{\pi}=$ 0.3 fm, $b=$ 0.6 fm, 
$z=$ 0.5). Data from P.Brauel et al.~\cite{b12} (modified): circles for 
$R\equiv\sigma_L/\sigma_T=$ 1. and squares for $R=$ 0.5;
$Q^2=$ 0.7 $GeV^2/c^2$ (a,c), $Q^2=$ 1.35 $GeV^2/c^2$ (b,d); 
$W=$ 1.9$\div$2.5 GeV.

{\bf Fig.11.} $Q^2$-dependence of the longitudinal kaon quasi-elastic knockout 
differential cross section $d\sigma_L/d\Omega$ at zero angle to the photon 
direction: (a) $p(e,e'K^+)\Lambda$, (b) $p(e,e'K^+)\Sigma^0$.
Solid lines: the scalar $^3P_0$ model ($b_{\pi}=$ 0.3, $b=$ 0.6 fm, $z=$ 
0.5). Data from Ref.~\cite{b24a}.

\newpage

%------------------------------------------------------------------------------
%------------------------------------------------------------------------------

%------------------------------------------------------------------------------
%------------------------------------------------------------------------------

%------------------------------------------------------------------------------
%------------------------------------------------------------------------------

%------------------------------------------------------------------------------
%------------------------------------------------------------------------------

%------------------------------------------------------------------------------
%------------------------------------------------------------------------------

%------------------------------------------------------------------------------
%------------------------------------------------------------------------------

%------------------------------------------------------------------------------
%------------------------------------------------------------------------------

%------------------------------------------------------------------------------

%------------------------------------------------------------------------------

%------------------------------------------------------------------------------

%------------------------------------------------------------------------------

%------------------------------------------------------------------------------
%------------------------------------------------------------------------------
%------------------------------------------------------------------------------
%%**************************************************
\end{document}